\documentclass[pdflatex,sn-mathphys-num]{sn-jnl}
\usepackage[T1]{fontenc}
\usepackage{graphicx}
\usepackage{multirow}
\usepackage{amsmath,amssymb,amsfonts}
\usepackage{amsthm}
\usepackage{mathrsfs}
\usepackage[title]{appendix}
\usepackage{xcolor}
\usepackage{textcomp}
\usepackage{manyfoot}
\usepackage{booktabs}
\usepackage{algorithm}
\usepackage{algorithmicx}
\usepackage{algpseudocode}
\usepackage{listings}
\theoremstyle{thmstyleone}

\theoremstyle{thmstyletwo}

\theoremstyle{thmstylethree}

\begin{document}
	
	\title[Article Title]{Phase Topology Stability of an Optical Vortex via an Electrically Controlled Twist-Planar Oriented Liquid Crystal Fresnel Lens}
	
	\author[1]{\fnm{Elena} \sur{Melnikova}}\email{Melnikova@bsu.by}
	\author[1]{\fnm{Katsiaryna} \sur{Pantsialeyeva}}\email{pantsialeyevakate@gmail.com}
	\author[1]{\fnm{Dmitry} \sur{Gorbach}}\email{gorbachdv@yandex.ru}
	\author[1]{\fnm{Alexei} \sur{Tolstik}}\email{Tolstik@bsu.by}
	\author[2]{\fnm{Sergei} \sur{Slussarenko Jr.}}\email{s.slussarenko@griffith.edu.au}
	\author*[3,4]{\fnm{Alina} \sur{Karabchevsky}}\email{a.karabchevsky@lancaster.ac.uk}
	\affil[1]{\orgdiv{Department of Laser Physics and Spectroscopy}, \orgname{Belarusian State University},\orgaddress{ \city{Minsk}, \country{Belarus}}}
	\affil[2]{\orgname{Griffith University}, \orgaddress{ \city{Brisbane}, \country{Australia}}}
	\affil*[3]{\orgdiv{School of Electrical and Computer Engineering}, \orgname{Ben-Gurion University of the Negev}, \orgaddress{\city{Beer-Sheva}, \postcode{8410501}, \country{Israel}}}
	\affil*[4]{\orgdiv{Department of Physics}, \orgname{Lancaster University}, \orgaddress{\city{ Lancaster}, \postcode{LA1 4YB}, \country{United Kingdom}}}
	
	\abstract{Optical vortices (OVs) have emerged as a revolutionary concept in modern photonics, offering a unique method of manipulating light beyond conventional Gaussian beams. Despite their vast potential, phase topology stability remains unaddressed, limiting their widespread adoption and performance in real-world environments. Here, we reveal the missing link to assessing the stability of optical vortices using an electrically tunable twist-planar liquid crystal (LC) Fresnel lens. The proposed LC-based lens leverages the birefringence and voltage-controlled reconfigurability of liquid crystals to dynamically probe the phase topology of singular beams. By modulating the LC orientation with an applied voltage, we restructure the optical phase in real-time without requiring modifications to the optical setup. The 3V and 35V voltage supply allows for the switch between the "topological charge detection" and "optical singular beam propagation" modes. This eliminates the need for additional optical elements, significantly simplifying the detection and characterization of vortex beams. 
		Experimental and theoretical investigations demonstrate that the vortex topology can be unambiguously identified from the intensity profile observed in the Fourier plane of a lens. Furthermore, the designed device features low power consumption, compact form factor, and seamless integration potential, making it a promising candidate for scalable vortex-based photonic systems.}
	
	
	\maketitle
	
	\section*{Main}\label{sec1}
	
	Since their discovery, optical vortices (OVs) — phase singular beams characterized by helical phase front and carrying orbital angular momentum (OAM), provide transformative applications in optical communication, quantum information processing, high-resolution imaging, and laser-material interactions \cite{padgett2017orbital, shen2019optical, huang2025integrated, session2025optical, hu2025topological, hu2025generalized, chen2024integrated}.  A defining feature of these beams is the presence of a phase singularity. At this point, the optical phase remains undefined, encircled by a helical phase structure that varies by 2$\pi\ell$, where $\ell$ denotes the topological charge of the vortex. This helical phase profile imparts an intrinsic OAM to the beam, a property that has led to breakthroughs in optical communication \cite{krenn2014communication, richardson2013space}, optical manipulation of microobjects, and the study of rheological properties through angular momentum transfer \cite{willner2015optical, bruce2021initiating}. Optical vortices also play a pivotal role in astronomical imaging \cite{aleksanyan2017multiple, aleksanyan2018high}, enabling high-contrast observations of exoplanets.
	
	Another fundamental characteristic of vortex beams is the presence of a central intensity null due to the phase singularity. This property is exploited in applications such as solar coronagraphy \cite{foo2005optical, serabyn2010image}, high-precision laser material processing \cite{masuda2017azo, takahashi2016picosecond}, optical trapping and controlled micro-particle motion \cite{otsu2014direct}, and ultra-high-resolution microscopy \cite{tan2010high, zhang2016perfect}.
	
	Various techniques have been developed to generate optical vortices, each leveraging different physical mechanisms. These include astigmatic mode conversion \cite{nye1974dislocations}, spiral phase plates \cite{beijersbergen1994helical, sueda2004laguerre}, q-plates \cite{marrucci2013q, melnikova2023achromatic, marrucci2012spin, kobashi2017broadband} and dynamically programmable wavefront shaping using spatial light modulators \cite{ostrovsky2013generation} and digital micromirror devices \cite{mirhosseini2013rapid}, photonic disclination cavities \cite{hwang2024vortex}. In the early 1990s, researchers began investigating the use of diffractive optical elements to generate optical vortices from beams with uniform phase fronts \cite{bazhenov1990laser, heckenberg1992generation}. A seminal study by Soskin, Vasnetsov, and Bazhenov demonstrated that introducing an optical dislocation into the grating structure—forming what are now known as fork or branched gratings—leads to the emergence of phase singularities in the diffracted beam \cite{bazhenov1990laser}. This foundational work remains a cornerstone in the field of singular optics.
	
	However, in practical systems, the stability of high-order optical vortices ($\mid\ell\mid$ $>$ 1) is fundamentally limited by imperfections in vortex generation techniques and the unavoidable presence of coherent background fields due to transmission, scattering, and reflection \cite{nye1974dislocations, basistiy1993optics, berry2001knotted}. These perturbations introduce instability in the phase singularities of high-order vortices, leading to their fragmentation into multiple first-order vortices ($\mid\ell\mid$ = 1) \cite{basistiy1993optics, ricci2012instability}. This effect manifests as a transformation of the vortex intensity distribution, with the appearance of multiple zero-intensity regions in the far field. While the singularity points may remain coalesced in the near field, particularly within the Rayleigh range, the presence of even a weak coherent background field in the far field induces significant separation of these singularities, with the degree of fragmentation increasing for higher topological charges \cite{basistiy1993optics, ricci2012instability}.
	
	Here, we exploit the inherent instability as a means of directly determining the topological charge of optical vortices. By applying a Fourier transform via a focusing lens, we transfer the field’s intensity distribution from the Fraunhofer region to the Fourier plane, greatly simplifying the experimental setup.
	
	Several established methods exist for measuring the topological charge of optical vortices, including astigmatic transformations via tilted or cylindrical lenses \cite{abramochkin1991beam, vaity2013measuring, denisenko2009determination}, spatial light modulators \cite{forbes2016creation}, diffraction through gratings \cite{kotlyar1998light, dai2015measuring} or structured apertures \cite{guo2009measuring, anderson2012measuring}, as well as interference-based techniques \cite{leach2002measuring, frkaczek2006new}. While effective, these approaches typically require the use of additional optical components in the beam path, complicating experimental setups and limiting real-time applications where simultaneous vortex generation and measurement are required.
	
	Here, we propose a novel approach that leverages the intrinsic splitting effect of optical vortices for a simple and direct analysis of their phase topology. Our method, supported by both experimental results and computational modeling, enables the unambiguous determination of the magnitude and sign of the topological charge without the need for additional optical elements. This is achieved by using an electrically switchable nematic liquid crystal (NLC) Fresnel lens, which we have developed. This lens's tunability and dynamic reconfigurability offer a compact, low-power, and scalable solution for real-time vortex characterization, with potential applications in optical communication, laser beam shaping and structured light applications, quantum photonic technologies, cryptography, and quantum computing.
	
	\section*{Results}\label{sec2}

	\subsection*{The Electrically Tunable NLC Fresnel Lens}
	
\begin{figure}[H]
		\centering
		\includegraphics[width=\linewidth]{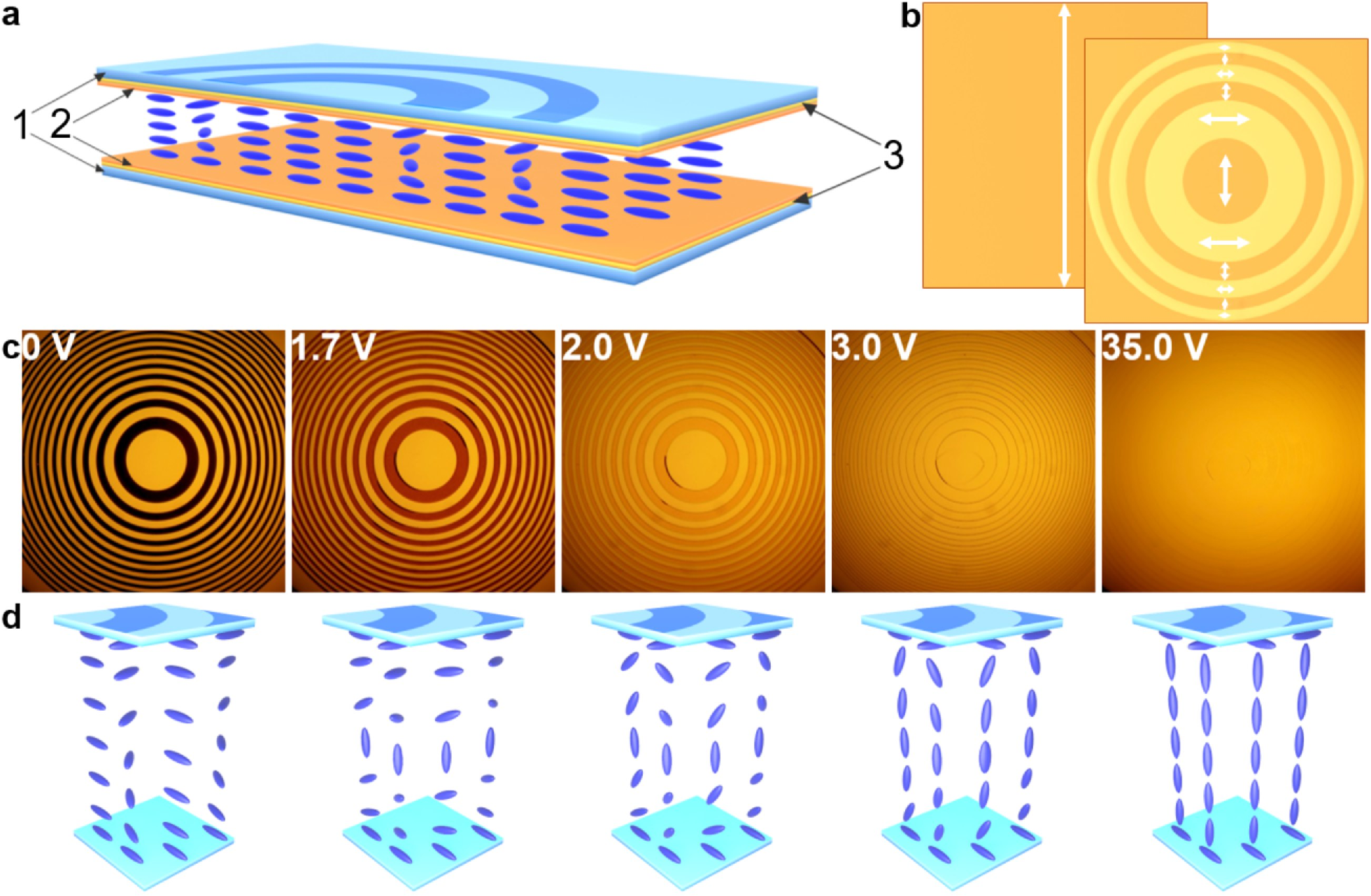}
		\caption{$\mid$ \textbf{Scheme of the NLC Fresnel lens.} \textbf{a}, Schematic diagram of the element with the orientation distribution of the liquid crystal director in the twist-planar oriented domains. \textbf{b}, Design of boundary conditions for the director of an NLC layer. The arrows schematically indicate the liquid crystal director's directions. \textbf{c}, Experimental micrographs of the electrically controlled NLC Fresnel lens placed between parallel polarizer and analyzer at different control voltage values \textit{U}: from 0 V to 35 V (from left to right) and \textbf{d}, the corresponding distribution of NLC molecules.}
		\label{fig:Schematic diagram of the element}
	\end{figure}

	A design of the proposed nematic liquid crystal Fresnel lens (Figure \ref{fig:Schematic diagram of the element}) is based on the domain orientation of the liquid crystal director. To set the initial orientation topology of the LC director (refractive index modulation), a photosensitive film \textit{3} of AtA-2 azo dye (20-30 nm thick) with an absorption band of 450-520 nm was deposited to both glass substrates \textit{1}, pre-coated with a layer of conductive indium tin oxide (ITO) \textit{2} for electrical control of the optical properties of the element \cite{mikulich2016influence, muravsky2020photoinduced}. Further, the orienting films on both substrates were exposed to linearly polarized radiation with a wavelength of 450 nm to set boundary conditions for a uniform planar orientation of the NLC director. Due to the reversibility of the orienting properties of AtA-2 azo dye, it is possible to re-irradiate the substrate with radiation with the direction of polarization rotated 90 degrees relative to the initial direction to form a domain structure. To create a structured element, re-exposure was carried out through an amplitude photomask, which is a Fresnel zone plate (15 pairs of concentric rings) with transparent even zones (the radius of the first transparent ring is 473 $\mu$m) and opaque odd zones (the radius of the first opaque ring is 333 $\mu$m). As a result, boundary conditions were set on the surfaces of the azo dye films (Figure \ref{fig:Schematic diagram of the element}b) for the formation of domains with a twist (the director rotates 90 degrees in the NLC layer) and planar (the director orientation is parallel and coincides on both substrates) director orientations in the nematic liquid crystal layer. The element was filled with a nematic liquid crystal in an isotropic phase after gluing the prepared substrates using fiber spacers with a diameter of 20 $\mu$m along the element's edges. Using the technology described above and in Supplementary Note 1, a diffraction grating with planar and twisted director orientation topologies was produced, which is an electrically switchable Fresnel NLC lens. 
	
	Previously, the twist-planar topology of the orientation of the NLC director was set using films of polymers (with absorption bands in ultraviolet) or azo dyes (sensitive in the blue region of the spectrum) \cite{kazak2010controlling, chigrinov2020photoaligning, wang2013switchable}. As practice shows, using photoorientants with an absorption band in the blue region of the spectrum significantly simplifies the process of manufacturing NLC elements. In the current work, the activating dose of AtA-2 azo dye was 0.3 J cm$^{-2}$, which is an order of magnitude less than the activation dose (5 J cm$^{-2}$) of polyimide dye PI-3744  in the work \cite{wang2013switchable}.
	
	The pronounced electro-optical response (modulation of the anisotropy of the refractive index \textit{$\Delta$n(U)}) of a nematic liquid crystal makes it possible to effectively control the properties of the manufactured structure using an external electric voltage applied to the electrodes of the cell (Figure \ref{fig:Schematic diagram of the element}c, d). In the voltage range from 0 V to 1.5 V, the twist-planar structure consists of two independent thin amplitude grids of rectangular stroke profile \cite{melnikova2022polarization, wkeglowski2017electro}. Twist-oriented LC domains rotate the plane of polarization of transmitted radiation by an angle of 90 degrees. The study of a structured lens by polarizing microscopy in the case when the lens was placed between a parallel polarizer and analyzer showed that areas with a twisted structure correspond to dark areas, and areas with a uniform planar orientation correspond to light areas. The supply of an external control voltage to the electrodes of the NLC element above the threshold value ($U_{th}=1.1$ V) causes an electro-optical response, which is expressed in the aspiration of the molecules to reorient along the lines of force of the electric field (Fredericks transition \cite{zhou2015optical, lucchetti2014light}). When the voltage reaches a value corresponding to the optical threshold ($U_{op}=1.5$ V), the Mogen condition \cite{mauguin1911cristaux} is disrupted in domains with a twist-oriented director due to the formation of a region with homeotropically arranged molecules and an indefinite azimuthal angle in the center of the LC layer parallel to the glass substrates. The reorientation leads to the transformation of amplitude gratings into a phase sinusoidal diffraction structure with a maximum value of diffraction efficiency (30\% \cite{melnikova2022polarization}) at a control voltage value of the order of 3 V. When the voltage reaches 35 V, the diffraction structure unwinds and the domain structure disappears (all NLC layer molecules are homeotropically oriented) \cite{mel2024electrically}.
	
	\subsection*{Determinaton of the Phase Topology of a Singular Beam}\label{subsec2}
	
	To study the phase topology of optical vortices after passing through a twist-planarly oriented NLC Fresnel lens, we constructed an experimental setup shown in Figure \ref{fig:Scheme}b and described in detail in Supplementary Note 2. The set-up consists of a radiation source He-Ne laser wavelength of 632.8 nm, collimating system which consist of an objective 20X and lens with focusing distance of 20cm to form a phase singular beam using a phase plate. The formed optical vortex with a topological charge $\ell$ evolves through the focusing NLC Fresnel lens (the optical axis of the singular beam passes through the center of the Fresnel lens). It is recorded by the charge-coupled device (CCD) behind the NLC element. Patterns of intensity distribution profiles were then imaged with the CCD camera and analysed on the computer.
	
	\begin{figure}[H]
		\centering
		\includegraphics[width=\linewidth]{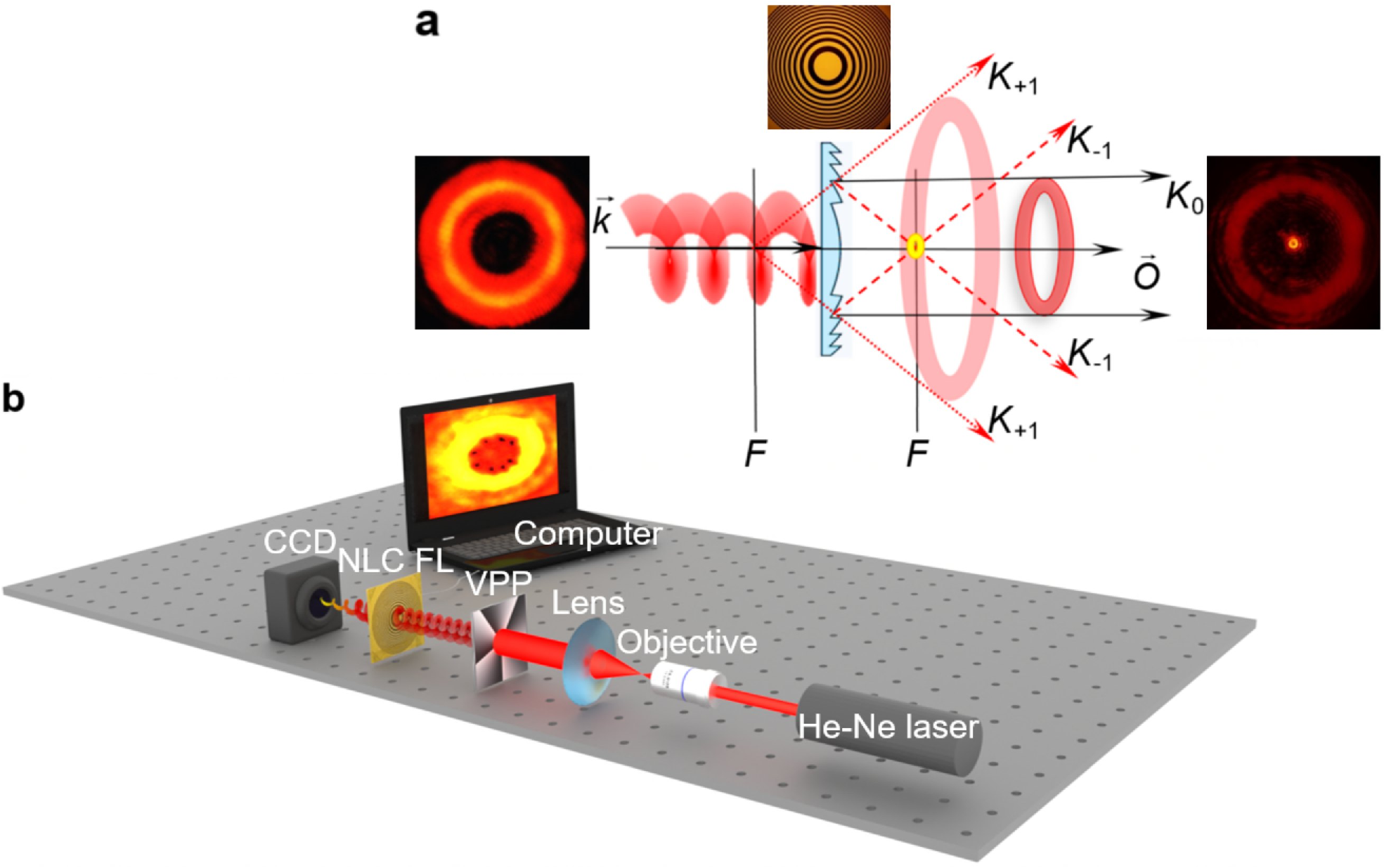}
		\caption{$\mid$ \textbf{Schematics to analyze singular beams.} \textbf{a}, Analysis and principle of determination of the phase topology and instability of optical vortices with NLC twist-planar Fresnel lens. \textbf{b}, The experimental setup to study the distribution profile of an optical vortex, with a He-Ne laser $\lambda=632.8$ nm, a 20X objective, a spherical lens, a vortex phase plate (VPP), a NLC Fresnel lens (NLC FL), a CCD camera, a computer.}
		\label{fig:Scheme}
	\end{figure}
	
\begin{figure}[H]
		\centering
		\includegraphics[width=\linewidth]{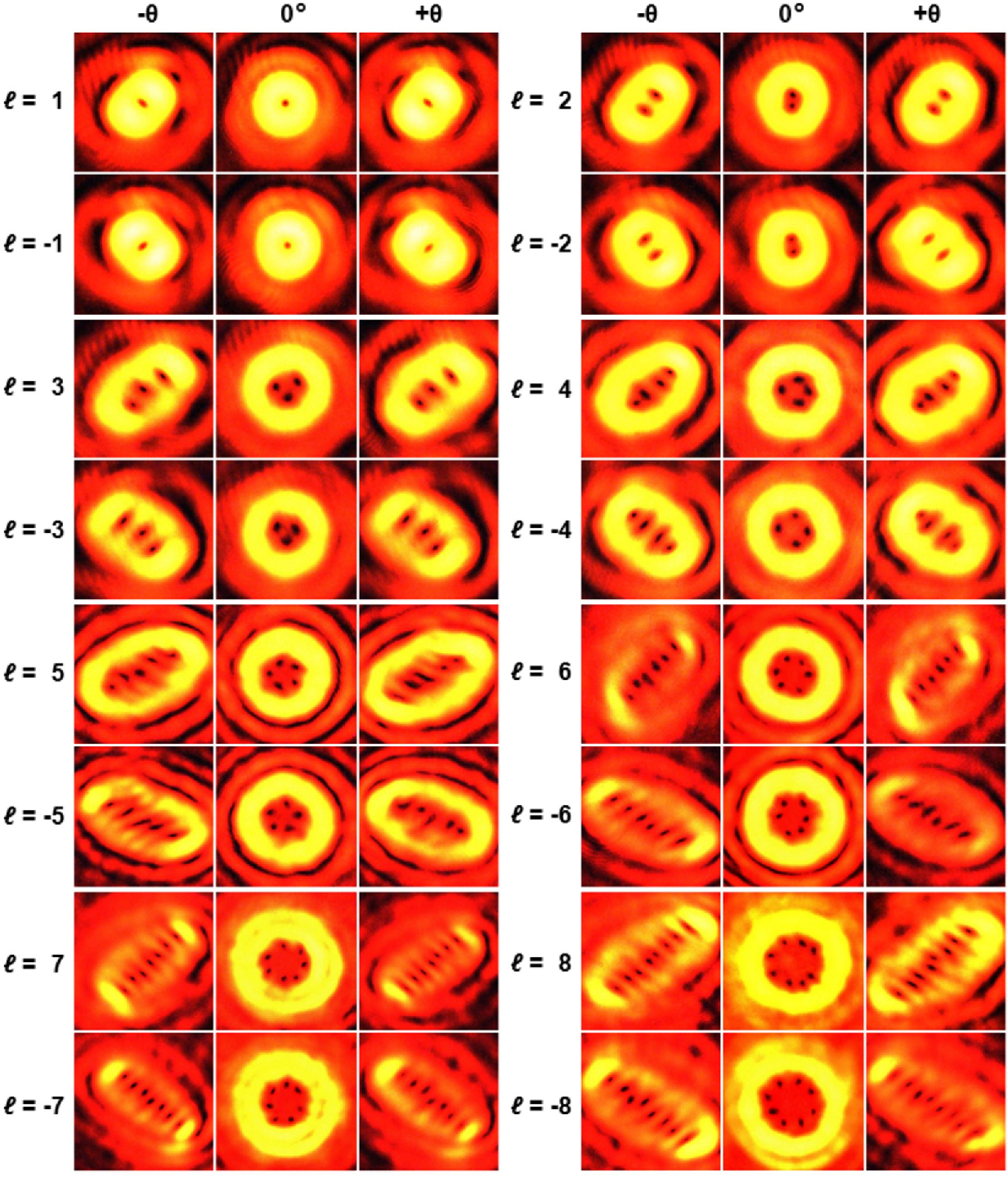}
		\caption{$\mid$ \textbf{Experimental characterization of the phase topology of OVs.} The distribution profile of the light field of a phase singular beam with topological charges $\ell$ from $\pm1$ to $\pm8$ in the focal plane of the Fresnel lens: under normal incidence of an optical vortex on the NLC lens ($\theta$ = 0 degrees) and when the radiation incidence at angles of $-\theta$ and $+\theta$ ($\theta$ increased from 9 to 18 degrees with increasing topological charge) to the normal of the lens.}
		\label{fig:All}
	\end{figure}

	Figure \ref{fig:All} shows experimental evidence of the normal centrosymmetric passage of an optical vortex with an integer topological charge through a nematic liquid crystal twist-planar oriented Fresnel lens in the focal plane of the element. A $\ell$-multiply degenerate intensity zero decays into $\ell$ isolated first-order intensity zeros located in the central region of the phase singular beam (Figure \ref{fig:All}, $\theta$ = 0 degrees). When the Fresnel lens is rotated around the vertical axis (changing the angle of incidence of radiation), an astigmatic transformation of the optical vortex occurs and $\ell$ isolated intensity zeros are observed (Figure \ref{fig:All}, $\pm\theta$) lying on a straight line in the transverse plane at an angle of $\pm45$ degrees to the plane of astigmatism, depending on the sign of the topological charge $\ell$ of the initial optical vortex. A positive charge ($\ell>0$) corresponds to an angle of +45 degrees, and a negative charge ($\ell<0$) corresponds to an angle of -45 degrees. Phase singular beams with topological charges $\ell$ from $\pm1$ to $\pm8$ have been experimentally analyzed. Experimental results and numerical simulation when an external control voltage is applied to a Fresnel NLC lens are shown in Supplementary Notes 2 and 4, accordingly.
	
\begin{figure}[H]
		\centering
		\includegraphics[width=\linewidth]{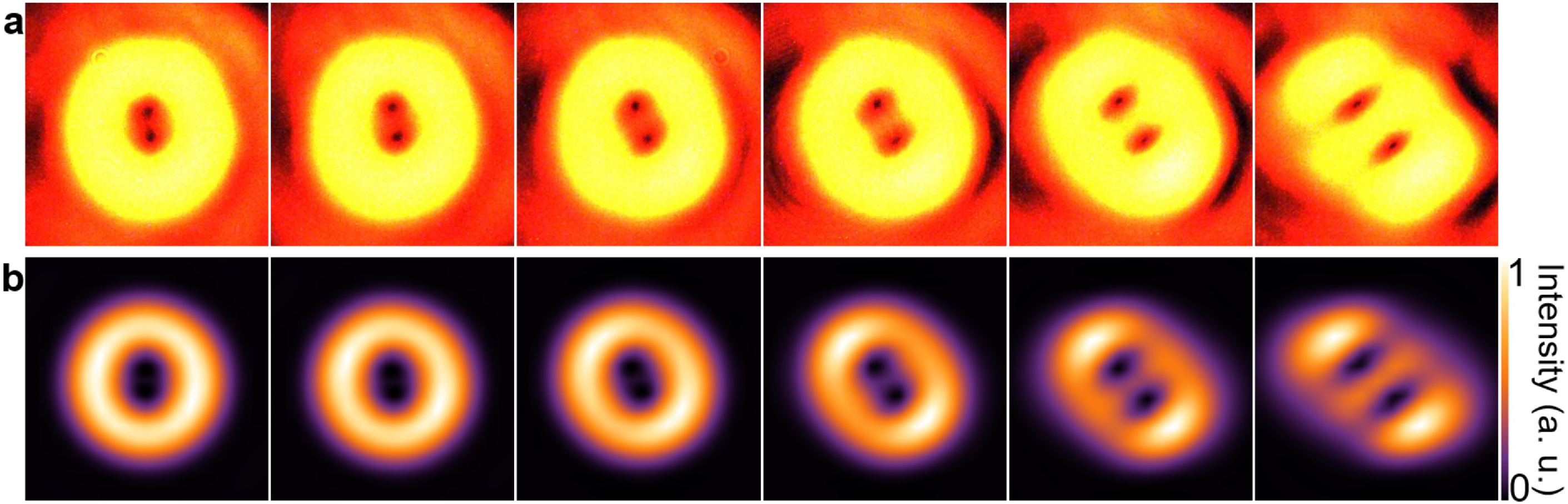}
		\caption{$\mid$ \textbf{Experimental and numerical research of the Phase Topology of OVs.} Experimental (\textbf{a}) and theoretical (\textbf{b}) results of the light field distribution profile of a phase singular beam with a topological charge $\ell$ = -2 in the focal plane of the Fresnel lens when the angle of incidence $\theta$ of radiation on the lens changes: from 0 to +10 degrees in increments of 2 degrees.}
		\label{fig:Angle2}
	\end{figure}

	For a beam with a topological charge of $\ell$ = -2, the intensity distributions in the focal plane when slightly changing the angle of incidence of radiation on the NLC lens from 0 (normal incidence) to 10 degrees in 2 degree increments (Figure \ref{fig:Angle2}) are analyzed. The obtained experimental results (Figure \ref{fig:Angle2}a) agree well with theoretical analyses (Figure \ref{fig:Angle2}b). The number of intensity dips is equal to the order (magnitude of the topological charge) of phase singular beam. Moreover, the sign of the charge is easily distinguished. As the angle of incidence of the studied radiation on the lens changes, the orientation of the light field distribution profile changes accordingly: modulated into elliptical and tilted distribution.
	
\begin{figure}[H]
		\centering
		\includegraphics[width=\linewidth]{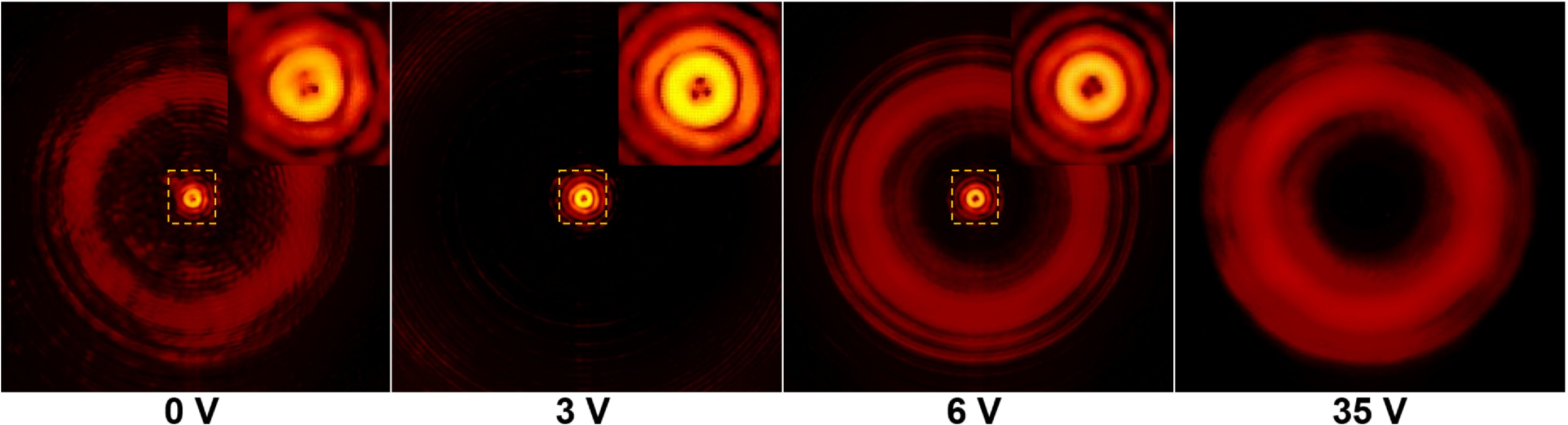}
		\caption{$\mid$ \textbf{Electrical switching of the NLC lens operation modes.} The distribution profile of the light field of the phase singular beam with a topological charge of $\mid\ell\mid$ = 3 in the focal plane of the Fresnel lens under normal incidence of the optical vortex on the NLC lens at different values of the control voltage (\textit{U}): from 0 V to 35 V (from left to right). The inserts show enlarged images of the selected areas of the photos.}
		\label{fig:focal plane}
	\end{figure}

	Figure \ref{fig:focal plane} shows the experimental results of a study of focusing a phase singular beam with a topological charge of $\mid\ell\mid$ = 3 by the NLC Fresnel lens at different values of the external control voltage applied to the lens. An analysis of the distribution profile of the optical vortex light field in the focal plane of the NLC lens showed that when a phase singular beam is diffracted on the NLC Fresnel lens, an optical vortex is observed in the direction of 0 diffraction order (direct radiation – Figure \ref{fig:Scheme}a, $K_0$) and radiation in the direction of -1 diffraction order (converging wave – Figure \ref{fig:Scheme}a, $K_{-1}$). The nonmonotonic dependence of the focusing properties of the manufactured element is shown. The diffraction efficiency is maximal at a voltage of about 3 V (the optimal voltage value \cite{mel2024electrically}) and we observe the absence of radiation in the direction of the diffraction order 0 ($K_0$). With a further increase of the value of the external voltage on the element, a monotonous decrease in diffraction efficiency is observed, due to the reorientation of liquid crystal molecules along the lines of electric field strength. Thus, the supply of the external control voltage of the order of 35 V (turning off the diffraction mode of operation of the NLC lens) makes it possible to realize the transmission mode of the incoming optical signal without removing the NLC Fresnel lens from the circuit.
	
\begin{figure}[H]
		\centering
		\includegraphics[width=\linewidth]{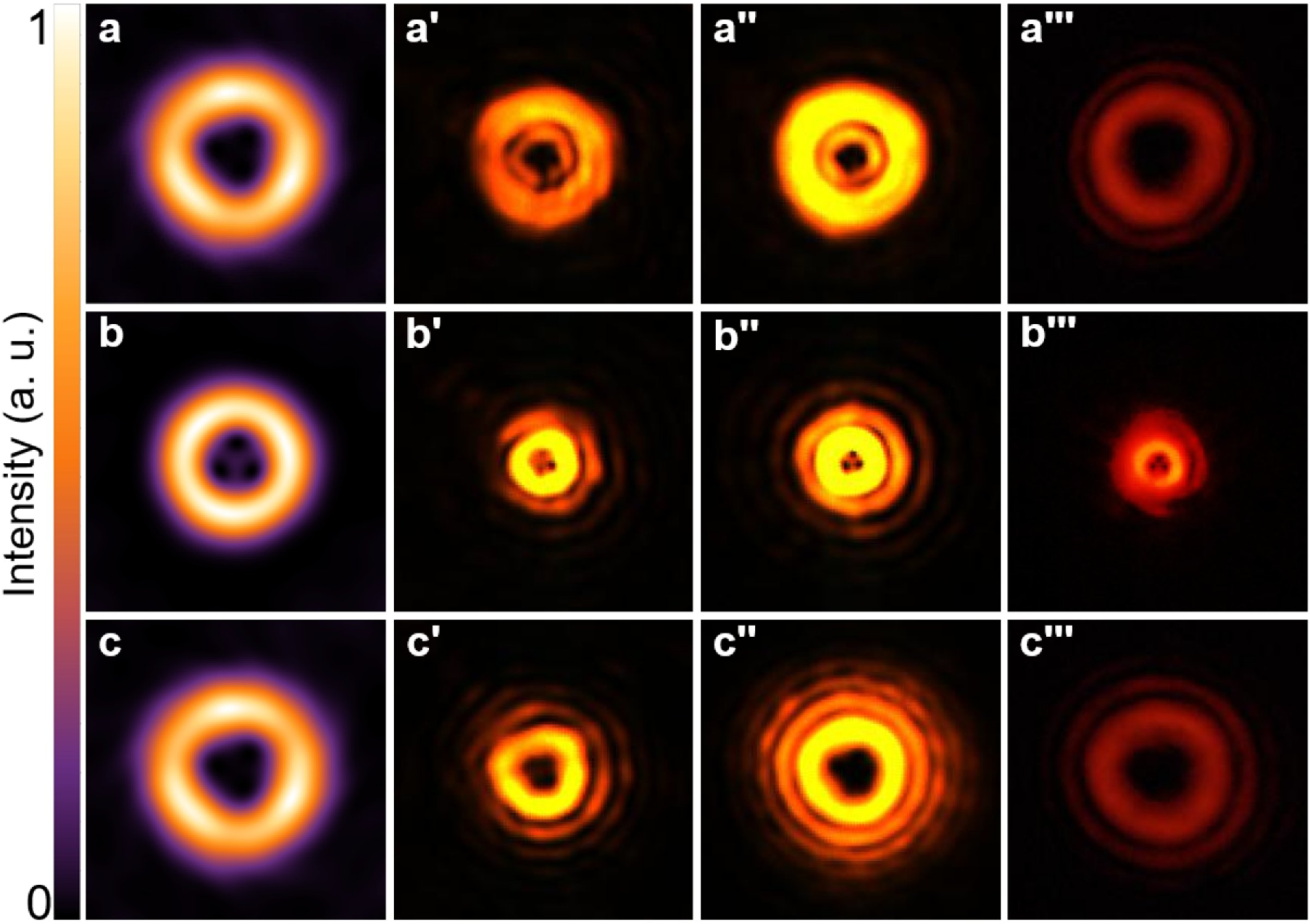}
		\caption{$\mid$ \textbf{Experimental and numerical investigation of the phase topology of OVs near the Fourier plane.} Light field distribution profile of a phase singular beam with a topological charge $\ell$=+3: 2 cm before (\textbf{a-a$^{\prime\prime\prime}$}), in (\textbf{b-b$^{\prime\prime\prime}$}) and 2 cm after (\textbf{c-c$^{\prime\prime\prime}$}) the focal plane of the lens. \textbf{a-c, a$^{\prime}$-c$^{\prime}$},  corresponds to use of Fresnel lens with control voltage \textit{U} = 0 V and \textbf{ a$^{\prime\prime}$-c$^{\prime\prime}$}, with \textit{U} = 3 V.\textbf{ a$^{\prime\prime\prime}$-c$^{\prime\prime\prime}$}, corresponds to use spherical lens. }
		\label{fig:focal plane2}
	\end{figure}

	Figure \ref{fig:focal plane2} illustrates the theoretical (Figure \ref{fig:focal plane2}a-c) and experimental (Figure \ref{fig:focal plane2}a$^{\prime}$-a$^{\prime\prime\prime}$, b$^{\prime}$-b$^{\prime\prime\prime}$, c$^{\prime}$-c$^{\prime\prime\prime}$) results of analyzing the distribution profile of the converging wave in the -1 diffraction order direction as radiation propagates: before focus (Figure \ref{fig:focal plane2}a-a$^{\prime\prime\prime}$), in focus (Figure \ref{fig:focal plane2}b-b$^{\prime\prime\prime}$) and after focus (Figure \ref{fig:focal plane2}c-c$^{\prime\prime\prime}$). It is shown that an optical vortex with a charge of $\ell$ = +3 is unstable in real conditions, as a result of which it splits into 3 separate vortices in the focal plane of the lens (Figure \ref{fig:focal plane2}b$^{\prime}$-b$^{\prime\prime\prime}$). The experiment was also performed with a spherical lens with the same focal length. The results obtained with its help (Figure \ref{fig:focal plane2}a$^{\prime\prime\prime}$-c$^{\prime\prime\prime}$) are qualitatively identical to the results obtained using the Fresnel lens (Figure \ref{fig:focal plane2}a$^{\prime}$-c$^{\prime}$, a$^{\prime\prime}$-c$^{\prime\prime}$). The study of the phase topology of optical vortices by coherent addition with spherical and plane waves is presented as Supplementary Note 3.
	
	\section*{Discussion}\label{sec13}
	
	Optical vortices hold immense promise for next-generation photonic technologies, offering new avenues for communication, computation, and material processing \cite{padgett2017orbital, shen2019optical}. However, their full potential remains untapped due to persistent technical challenges. As is known, the presence of noise in a singular light field leads to the decay of high-order optical vortices into several vortices with a charge $\mid\ell\mid$ = 1, for example, in the presence of a coherent background \cite{soskin1997topological}. Thus, under real conditions, Laguerre-Gaussian beams with a topological charge 
	$\mid\ell\mid$ $>$ 1 persist over limited distances and disintegrate in the presence of noise. The decay of a high-order singular phase beam into optical vortices with single charges $\mid\ell\mid$ = 1 \cite{basistiy1993optics, ricci2012instability} with the preservation of the total topological charge is due to the presence of noise caused by the imperfection of optical elements with which the corresponding phase transformation is carried out (optical phase plates with smoothly variable thickness, spatial LC light modulators, holographic elements, q-plates, etc.). 
	
	The main non-resonator elements for vortex formation are phase isotropic elements with spirally varying material thickness \cite{forbes2016creation, ma2013study,szatkowski2020generation} and q-plates - anisotropic centrally symmetric azimuthally structured optical elements \cite{nersisyan2009fabrication, marrucci2012spin, slussarenko2011tunable}. The fabrication of such elements requires overcoming the problems of design accuracy and technological limitations associated with the high precision of positioning the center of the element (with an accuracy of less than 1 $\mu$m) around which the optical properties change.
	
	Another technology for creating phase singular beams is the use of spatial liquid crystal light modulators \cite{forbes2016creation, ma2013study,szatkowski2020generation}. However, the limitation of spatial resolution (for example, the pixel size of 3.74 $\mu$m (EXULUS-4K1, Thorlabs)), phase shift discretization, and the presence of pixel spacing lead to a deterioration in the quality of the generated beam.
	
	The technological and constructive disadvantages of the described methods of forming singular phase fields lead to the fact that in realistic scenarios, vortexes are unstable and disintegrate when propagating in free space. A CCD camera is usually used to analyze the evolution of vortices in an experiment, moving it along the beam propagation axis to control the intensity distribution in its cross section. However, the distance to which the camera can be moved is limited by the possibility of experimental conditions. In this regard, the qualitative analysis of the far zone for assessing the stability of optical vortices becomes a rather serious task that needs to be solved. Addressing the unmet needs of phase topology stability is critical for realizing the transformative impact of optical vortices in both fundamental research and real-world applications. As a solution, we propose using a focusing lens to transfer the intensity distribution of the light field from the Fraunhofer zone to the Fourier zone. The Fourier transform using a lens allows for an effective analysis of the stability of phase singular beams (Figures \ref{fig:All}, \ref{fig:focal plane2}, Supplementary Note 4).
	
	Interference methods \cite{courtial1998measurement, harris1994laser, cui2019determining, denisenko2009determination} are traditionally used to determine the topological charge of optical vortices, requiring complex optical circuits with multiple elements to produce a coherent plane or spherical wave. However, the need for a coherent reference wave, the presence of aberrations, defects in optical elements, and alignment errors can significantly distort the interference pattern, make it difficult to determine the topological charge, and limit practical applications. This stimulated the development of alternative approaches to the characterization of optical vortices. In recent years, diffraction methods \cite{melo2018direct, han2011measuring, izdebskaya2012observation, mesquita2011engineering} for determining topological charge have been actively developing, which have a number of advantages over interference methods. They require fewer optical elements and are more resistant to external influences. At the same time, diffraction methods make it possible to carry out measurements in real time without the need for complex adjustment of the optical circuit and without using a coherent reference wave. However, despite the large number of methods already developed, they all have some limitations and difficulties in implementation. Therefore, at present, determining the phase topology of optical vortices is one of the key tasks of modern photonics to expand the field of application of phase singular beams.
	
	We propose using the effect of "splitting" an optical vortex with a charge $\ell$ in the far zone into $\mid\ell\mid$ separate regions of zero intensity as a new effective and simple method for determining the topological charge of a phase singular beam (Figure \ref{fig:All}). Hence, the generation of $\mid\ell\mid$ isolated intensity zeros in the focal plane of a spherical and the NLC Fresnel lenses when a phase singular beam with topological charge $\ell$ passes through it has been numerically and experimentally demonstrated  (Figures \ref{fig:All}, \ref{fig:focal plane2}, Supplementary Notes 2, 3, 4). The addition of astigmatism to the beam by rotating the lens leads to alignment of the formed intensity zeros along a straight line at an angle of $\pm45$ degrees to the plane of the introduced astigmatism  (Figures \ref{fig:All}, \ref{fig:Angle2}, Supplementary Notes 2, 4). When the topological charge is positive, the angle will be $+45$ degrees and when the charge is negative, it will be $-45$ degrees. We also suggest using the NLC Fresnel lens as a focusing element (Figure \ref{fig:Schematic diagram of the element}, Supplementary Note 1). The use of the LC lens allows for electrical switching of the "optical singular beam propagation" and "topological charge detection" modes in real time without changing the optical circuit  (Figure \ref{fig:focal plane}).
	
	\backmatter
	\bmhead{Acknowledgements}
	
	The research was supported by the Israel Science Foundation (ISF) - no. 1023/24 and the state program of scientific research of Belarus "Convergence-2025".

	\bibliographystyle{sn-mathphys-num}
	\bibliography{sn-bibliography}


\begin{thebibliography}{73}
\ifx \bisbn   \undefined \def \bisbn  #1{ISBN #1}\fi
\ifx \binits  \undefined \def \binits#1{#1}\fi
\ifx \bauthor  \undefined \def \bauthor#1{#1}\fi
\ifx \batitle  \undefined \def \batitle#1{#1}\fi
\ifx \bjtitle  \undefined \def \bjtitle#1{#1}\fi
\ifx \bvolume  \undefined \def \bvolume#1{\textbf{#1}}\fi
\ifx \byear  \undefined \def \byear#1{#1}\fi
\ifx \bissue  \undefined \def \bissue#1{#1}\fi
\ifx \bfpage  \undefined \def \bfpage#1{#1}\fi
\ifx \blpage  \undefined \def \blpage #1{#1}\fi
\ifx \burl  \undefined \def \burl#1{\textsf{#1}}\fi
\ifx \doiurl  \undefined \def \doiurl#1{\url{https://doi.org/#1}}\fi
\ifx \betal  \undefined \def \betal{\textit{et al.}}\fi
\ifx \binstitute  \undefined \def \binstitute#1{#1}\fi
\ifx \binstitutionaled  \undefined \def \binstitutionaled#1{#1}\fi
\ifx \bctitle  \undefined \def \bctitle#1{#1}\fi
\ifx \beditor  \undefined \def \beditor#1{#1}\fi
\ifx \bpublisher  \undefined \def \bpublisher#1{#1}\fi
\ifx \bbtitle  \undefined \def \bbtitle#1{#1}\fi
\ifx \bedition  \undefined \def \bedition#1{#1}\fi
\ifx \bseriesno  \undefined \def \bseriesno#1{#1}\fi
\ifx \blocation  \undefined \def \blocation#1{#1}\fi
\ifx \bsertitle  \undefined \def \bsertitle#1{#1}\fi
\ifx \bsnm \undefined \def \bsnm#1{#1}\fi
\ifx \bsuffix \undefined \def \bsuffix#1{#1}\fi
\ifx \bparticle \undefined \def \bparticle#1{#1}\fi
\ifx \barticle \undefined \def \barticle#1{#1}\fi
\bibcommenthead
\ifx \bconfdate \undefined \def \bconfdate #1{#1}\fi
\ifx \botherref \undefined \def \botherref #1{#1}\fi
\ifx \url \undefined \def \url#1{\textsf{#1}}\fi
\ifx \bchapter \undefined \def \bchapter#1{#1}\fi
\ifx \bbook \undefined \def \bbook#1{#1}\fi
\ifx \bcomment \undefined \def \bcomment#1{#1}\fi
\ifx \oauthor \undefined \def \oauthor#1{#1}\fi
\ifx \citeauthoryear \undefined \def \citeauthoryear#1{#1}\fi
\ifx \endbibitem  \undefined \def \endbibitem {}\fi
\ifx \bconflocation  \undefined \def \bconflocation#1{#1}\fi
\ifx \arxivurl  \undefined \def \arxivurl#1{\textsf{#1}}\fi
\csname PreBibitemsHook\endcsname

\bibitem[\protect\citeauthoryear{Padgett}{2017}]{padgett2017orbital}
\begin{barticle}
\bauthor{\bsnm{Padgett}, \binits{M.J.}}:
\batitle{Orbital angular momentum 25 years on}.
\bjtitle{Optics express}
\bvolume{25}(\bissue{10}),
\bfpage{11265}--\blpage{11274}
(\byear{2017})
\end{barticle}
\endbibitem

\bibitem[\protect\citeauthoryear{Shen et~al.}{2019}]{shen2019optical}
\begin{barticle}
\bauthor{\bsnm{Shen}, \binits{Y.}},
\bauthor{\bsnm{Wang}, \binits{X.}},
\bauthor{\bsnm{Xie}, \binits{Z.}},
\bauthor{\bsnm{Min}, \binits{C.}},
\bauthor{\bsnm{Fu}, \binits{X.}},
\bauthor{\bsnm{Liu}, \binits{Q.}},
\bauthor{\bsnm{Gong}, \binits{M.}},
\bauthor{\bsnm{Yuan}, \binits{X.}}:
\batitle{Optical vortices 30 years on: Oam manipulation from topological charge
  to multiple singularities}.
\bjtitle{Light: Science \& Applications}
\bvolume{8}(\bissue{1}),
\bfpage{90}
(\byear{2019})
\end{barticle}
\endbibitem

\bibitem[\protect\citeauthoryear{Huang et~al.}{2025}]{huang2025integrated}
\begin{botherref}
\oauthor{\bsnm{Huang}, \binits{J.}},
\oauthor{\bsnm{Mao}, \binits{J.}},
\oauthor{\bsnm{Li}, \binits{X.}},
\oauthor{\bsnm{Yuan}, \binits{J.}},
\oauthor{\bsnm{Zheng}, \binits{Y.}},
\oauthor{\bsnm{Zhai}, \binits{C.}},
\oauthor{\bsnm{Dai}, \binits{T.}},
\oauthor{\bsnm{Fu}, \binits{Z.}},
\oauthor{\bsnm{Bao}, \binits{J.}},
\oauthor{\bsnm{Yang}, \binits{Y.}}, et al.:
Integrated optical entangled quantum vortex emitters.
Nature Photonics,
1--8
(2025)
\end{botherref}
\endbibitem

\bibitem[\protect\citeauthoryear{Session et~al.}{2025}]{session2025optical}
\begin{botherref}
\oauthor{\bsnm{Session}, \binits{D.}},
\oauthor{\bsnm{Jalali~Mehrabad}, \binits{M.}},
\oauthor{\bsnm{Paithankar}, \binits{N.}},
\oauthor{\bsnm{Grass}, \binits{T.}},
\oauthor{\bsnm{Eckhardt}, \binits{C.J.}},
\oauthor{\bsnm{Cao}, \binits{B.}},
\oauthor{\bsnm{Gustavo Su{\'a}rez~Forero}, \binits{D.}},
\oauthor{\bsnm{Li}, \binits{K.}},
\oauthor{\bsnm{Alam}, \binits{M.S.}},
\oauthor{\bsnm{Watanabe}, \binits{K.}}, et al.:
Optical pumping of electronic quantum hall states with vortex light.
Nature Photonics,
1--6
(2025)
\end{botherref}
\endbibitem

\bibitem[\protect\citeauthoryear{Hu et~al.}{2025a}]{hu2025topological}
\begin{botherref}
\oauthor{\bsnm{Hu}, \binits{Z.}},
\oauthor{\bsnm{Bongiovanni}, \binits{D.}},
\oauthor{\bsnm{Wang}, \binits{Z.}},
\oauthor{\bsnm{Wang}, \binits{X.}},
\oauthor{\bsnm{Song}, \binits{D.}},
\oauthor{\bsnm{Xu}, \binits{J.}},
\oauthor{\bsnm{Morandotti}, \binits{R.}},
\oauthor{\bsnm{Buljan}, \binits{H.}},
\oauthor{\bsnm{Chen}, \binits{Z.}}:
Topological orbital angular momentum extraction and twofold protection of
  vortex transport.
Nature Photonics,
1--8
(2025)
\end{botherref}
\endbibitem

\bibitem[\protect\citeauthoryear{Hu et~al.}{2025b}]{hu2025generalized}
\begin{botherref}
\oauthor{\bsnm{Hu}, \binits{J.}},
\oauthor{\bsnm{Eriksson}, \binits{M.}},
\oauthor{\bsnm{Gigan}, \binits{S.}},
\oauthor{\bsnm{Fickler}, \binits{R.}}:
Generalized angle--orbital angular momentum talbot effect and modulo mode
  sorting.
Nature Photonics,
1--8
(2025)
\end{botherref}
\endbibitem

\bibitem[\protect\citeauthoryear{Chen et~al.}{2024}]{chen2024integrated}
\begin{barticle}
\bauthor{\bsnm{Chen}, \binits{B.}},
\bauthor{\bsnm{Zhou}, \binits{Y.}},
\bauthor{\bsnm{Liu}, \binits{Y.}},
\bauthor{\bsnm{Ye}, \binits{C.}},
\bauthor{\bsnm{Cao}, \binits{Q.}},
\bauthor{\bsnm{Huang}, \binits{P.}},
\bauthor{\bsnm{Kim}, \binits{C.}},
\bauthor{\bsnm{Zheng}, \binits{Y.}},
\bauthor{\bsnm{Oxenl{\o}we}, \binits{L.K.}},
\bauthor{\bsnm{Yvind}, \binits{K.}}, \betal:
\batitle{Integrated optical vortex microcomb}.
\bjtitle{Nature Photonics}
\bvolume{18}(\bissue{6}),
\bfpage{625}--\blpage{631}
(\byear{2024})
\end{barticle}
\endbibitem

\bibitem[\protect\citeauthoryear{Krenn et~al.}{2014}]{krenn2014communication}
\begin{barticle}
\bauthor{\bsnm{Krenn}, \binits{M.}},
\bauthor{\bsnm{Fickler}, \binits{R.}},
\bauthor{\bsnm{Fink}, \binits{M.}},
\bauthor{\bsnm{Handsteiner}, \binits{J.}},
\bauthor{\bsnm{Malik}, \binits{M.}},
\bauthor{\bsnm{Scheidl}, \binits{T.}},
\bauthor{\bsnm{Ursin}, \binits{R.}},
\bauthor{\bsnm{Zeilinger}, \binits{A.}}:
\batitle{Communication with spatially modulated light through turbulent air
  across vienna}.
\bjtitle{New Journal of Physics}
\bvolume{16}(\bissue{11}),
\bfpage{113028}
(\byear{2014})
\end{barticle}
\endbibitem

\bibitem[\protect\citeauthoryear{Richardson et~al.}{2013}]{richardson2013space}
\begin{barticle}
\bauthor{\bsnm{Richardson}, \binits{D.J.}},
\bauthor{\bsnm{Fini}, \binits{J.M.}},
\bauthor{\bsnm{Nelson}, \binits{L.E.}}:
\batitle{Space-division multiplexing in optical fibres}.
\bjtitle{Nature photonics}
\bvolume{7}(\bissue{5}),
\bfpage{354}--\blpage{362}
(\byear{2013})
\end{barticle}
\endbibitem

\bibitem[\protect\citeauthoryear{Willner et~al.}{2015}]{willner2015optical}
\begin{barticle}
\bauthor{\bsnm{Willner}, \binits{A.E.}},
\bauthor{\bsnm{Huang}, \binits{H.}},
\bauthor{\bsnm{Yan}, \binits{Y.}},
\bauthor{\bsnm{Ren}, \binits{Y.}},
\bauthor{\bsnm{Ahmed}, \binits{N.}},
\bauthor{\bsnm{Xie}, \binits{G.}},
\bauthor{\bsnm{Bao}, \binits{C.}},
\bauthor{\bsnm{Li}, \binits{L.}},
\bauthor{\bsnm{Cao}, \binits{Y.}},
\bauthor{\bsnm{Zhao}, \binits{Z.}}, \betal:
\batitle{Optical communications using orbital angular momentum beams}.
\bjtitle{Advances in optics and photonics}
\bvolume{7}(\bissue{1}),
\bfpage{66}--\blpage{106}
(\byear{2015})
\end{barticle}
\endbibitem

\bibitem[\protect\citeauthoryear{Bruce et~al.}{2021}]{bruce2021initiating}
\begin{barticle}
\bauthor{\bsnm{Bruce}, \binits{G.D.}},
\bauthor{\bsnm{Rodr{\'\i}guez-Sevilla}, \binits{P.}},
\bauthor{\bsnm{Dholakia}, \binits{K.}}:
\batitle{Initiating revolutions for optical manipulation: the origins and
  applications of rotational dynamics of trapped particles}.
\bjtitle{Advances in Physics: X}
\bvolume{6}(\bissue{1}),
\bfpage{1838322}
(\byear{2021})
\end{barticle}
\endbibitem

\bibitem[\protect\citeauthoryear{Aleksanyan
  et~al.}{2017}]{aleksanyan2017multiple}
\begin{barticle}
\bauthor{\bsnm{Aleksanyan}, \binits{A.}},
\bauthor{\bsnm{Kravets}, \binits{N.}},
\bauthor{\bsnm{Brasselet}, \binits{E.}}:
\batitle{Multiple-star system adaptive vortex coronagraphy using a liquid
  crystal light valve}.
\bjtitle{Physical review letters}
\bvolume{118}(\bissue{20}),
\bfpage{203902}
(\byear{2017})
\end{barticle}
\endbibitem

\bibitem[\protect\citeauthoryear{Aleksanyan and
  Brasselet}{2018}]{aleksanyan2018high}
\begin{barticle}
\bauthor{\bsnm{Aleksanyan}, \binits{A.}},
\bauthor{\bsnm{Brasselet}, \binits{E.}}:
\batitle{High-charge and multiple-star vortex coronagraphy from stacked vector
  vortex phase masks}.
\bjtitle{Optics Letters}
\bvolume{43}(\bissue{3}),
\bfpage{383}--\blpage{386}
(\byear{2018})
\end{barticle}
\endbibitem

\bibitem[\protect\citeauthoryear{Foo et~al.}{2005}]{foo2005optical}
\begin{barticle}
\bauthor{\bsnm{Foo}, \binits{G.}},
\bauthor{\bsnm{Palacios}, \binits{D.M.}},
\bauthor{\bsnm{Swartzlander~Jr}, \binits{G.A.}}:
\batitle{Optical vortex coronagraph}.
\bjtitle{Optics letters}
\bvolume{30}(\bissue{24}),
\bfpage{3308}--\blpage{3310}
(\byear{2005})
\end{barticle}
\endbibitem

\bibitem[\protect\citeauthoryear{Serabyn et~al.}{2010}]{serabyn2010image}
\begin{barticle}
\bauthor{\bsnm{Serabyn}, \binits{E.}},
\bauthor{\bsnm{Mawet}, \binits{D.}},
\bauthor{\bsnm{Burruss}, \binits{R.}}:
\batitle{An image of an exoplanet separated by two diffraction beamwidths from
  a star}.
\bjtitle{Nature}
\bvolume{464}(\bissue{7291}),
\bfpage{1018}--\blpage{1020}
(\byear{2010})
\end{barticle}
\endbibitem

\bibitem[\protect\citeauthoryear{Masuda et~al.}{2017}]{masuda2017azo}
\begin{barticle}
\bauthor{\bsnm{Masuda}, \binits{K.}},
\bauthor{\bsnm{Nakano}, \binits{S.}},
\bauthor{\bsnm{Barada}, \binits{D.}},
\bauthor{\bsnm{Kumakura}, \binits{M.}},
\bauthor{\bsnm{Miyamoto}, \binits{K.}},
\bauthor{\bsnm{Omatsu}, \binits{T.}}:
\batitle{Azo-polymer film twisted to form a helical surface relief by
  illumination with a circularly polarized gaussian beam}.
\bjtitle{Optics Express}
\bvolume{25}(\bissue{11}),
\bfpage{12499}--\blpage{12507}
(\byear{2017})
\end{barticle}
\endbibitem

\bibitem[\protect\citeauthoryear{Takahashi
  et~al.}{2016}]{takahashi2016picosecond}
\begin{barticle}
\bauthor{\bsnm{Takahashi}, \binits{F.}},
\bauthor{\bsnm{Miyamoto}, \binits{K.}},
\bauthor{\bsnm{Hidai}, \binits{H.}},
\bauthor{\bsnm{Yamane}, \binits{K.}},
\bauthor{\bsnm{Morita}, \binits{R.}},
\bauthor{\bsnm{Omatsu}, \binits{T.}}:
\batitle{Picosecond optical vortex pulse illumination forms a monocrystalline
  silicon needle}.
\bjtitle{Scientific reports}
\bvolume{6}(\bissue{1}),
\bfpage{21738}
(\byear{2016})
\end{barticle}
\endbibitem

\bibitem[\protect\citeauthoryear{Otsu et~al.}{2014}]{otsu2014direct}
\begin{barticle}
\bauthor{\bsnm{Otsu}, \binits{T.}},
\bauthor{\bsnm{Ando}, \binits{T.}},
\bauthor{\bsnm{Takiguchi}, \binits{Y.}},
\bauthor{\bsnm{Ohtake}, \binits{Y.}},
\bauthor{\bsnm{Toyoda}, \binits{H.}},
\bauthor{\bsnm{Itoh}, \binits{H.}}:
\batitle{Direct evidence for three-dimensional off-axis trapping with single
  laguerre-gaussian beam}.
\bjtitle{Scientific reports}
\bvolume{4}(\bissue{1}),
\bfpage{4579}
(\byear{2014})
\end{barticle}
\endbibitem

\bibitem[\protect\citeauthoryear{Tan et~al.}{2010}]{tan2010high}
\begin{botherref}
\oauthor{\bsnm{Tan}, \binits{P.}},
\oauthor{\bsnm{Yuan}, \binits{X.-C.}},
\oauthor{\bsnm{Yuan}, \binits{G.}},
\oauthor{\bsnm{Wang}, \binits{Q.}}:
High-resolution wide-field standing-wave surface plasmon resonance fluorescence
  microscopy with optical vortices.
Applied Physics Letters
\textbf{97}(24)
(2010)
\end{botherref}
\endbibitem

\bibitem[\protect\citeauthoryear{Zhang et~al.}{2016}]{zhang2016perfect}
\begin{botherref}
\oauthor{\bsnm{Zhang}, \binits{C.}},
\oauthor{\bsnm{Min}, \binits{C.}},
\oauthor{\bsnm{Du}, \binits{L.}},
\oauthor{\bsnm{Yuan}, \binits{X.-C.}}:
Perfect optical vortex enhanced surface plasmon excitation for plasmonic
  structured illumination microscopy imaging.
Applied Physics Letters
\textbf{108}(20)
(2016)
\end{botherref}
\endbibitem

\bibitem[\protect\citeauthoryear{Nye and Berry}{1974}]{nye1974dislocations}
\begin{barticle}
\bauthor{\bsnm{Nye}, \binits{J.F.}},
\bauthor{\bsnm{Berry}, \binits{M.V.}}:
\batitle{Dislocations in wave trains}.
\bjtitle{Proceedings of the Royal Society of London. A. Mathematical and
  Physical Sciences}
\bvolume{336}(\bissue{1605}),
\bfpage{165}--\blpage{190}
(\byear{1974})
\end{barticle}
\endbibitem

\bibitem[\protect\citeauthoryear{Beijersbergen
  et~al.}{1994}]{beijersbergen1994helical}
\begin{barticle}
\bauthor{\bsnm{Beijersbergen}, \binits{M.}},
\bauthor{\bsnm{Coerwinkel}, \binits{R.}},
\bauthor{\bsnm{Kristensen}, \binits{M.}},
\bauthor{\bsnm{Woerdman}, \binits{J.}}:
\batitle{Helical-wavefront laser beams produced with a spiral phaseplate}.
\bjtitle{Optics communications}
\bvolume{112}(\bissue{5-6}),
\bfpage{321}--\blpage{327}
(\byear{1994})
\end{barticle}
\endbibitem

\bibitem[\protect\citeauthoryear{Sueda et~al.}{2004}]{sueda2004laguerre}
\begin{barticle}
\bauthor{\bsnm{Sueda}, \binits{K.}},
\bauthor{\bsnm{Miyaji}, \binits{G.}},
\bauthor{\bsnm{Miyanaga}, \binits{N.}},
\bauthor{\bsnm{Nakatsuka}, \binits{M.}}:
\batitle{Laguerre-gaussian beam generated with a multilevel spiral phase plate
  for high intensity laser pulses}.
\bjtitle{Optics express}
\bvolume{12}(\bissue{15}),
\bfpage{3548}--\blpage{3553}
(\byear{2004})
\end{barticle}
\endbibitem

\bibitem[\protect\citeauthoryear{Marrucci}{2013}]{marrucci2013q}
\begin{barticle}
\bauthor{\bsnm{Marrucci}, \binits{L.}}:
\batitle{The q-plate and its future}.
\bjtitle{Journal of Nanophotonics}
\bvolume{7}(\bissue{1}),
\bfpage{078598}--\blpage{078598}
(\byear{2013})
\end{barticle}
\endbibitem

\bibitem[\protect\citeauthoryear{Melnikova
  et~al.}{2023}]{melnikova2023achromatic}
\begin{barticle}
\bauthor{\bsnm{Melnikova}, \binits{E.}},
\bauthor{\bsnm{Tolstik}, \binits{A.}},
\bauthor{\bsnm{Gorbach}, \binits{D.}},
\bauthor{\bsnm{Stanevich}, \binits{V.Y.}},
\bauthor{\bsnm{Kukhta}, \binits{I.}},
\bauthor{\bsnm{Chepeleva}, \binits{D.}},
\bauthor{\bsnm{Murauski}, \binits{A.A.}},
\bauthor{\bsnm{Muravsky}, \binits{A.A.}}:
\batitle{Achromatic switchable liquid-crystal twist-q-plate}.
\bjtitle{Journal of Applied Spectroscopy}
\bvolume{90}(\bissue{2}),
\bfpage{427}--\blpage{435}
(\byear{2023})
\end{barticle}
\endbibitem

\bibitem[\protect\citeauthoryear{Marrucci et~al.}{2012}]{marrucci2012spin}
\begin{barticle}
\bauthor{\bsnm{Marrucci}, \binits{L.}},
\bauthor{\bsnm{Karimi}, \binits{E.}},
\bauthor{\bsnm{Slussarenko}, \binits{S.}},
\bauthor{\bsnm{Piccirillo}, \binits{B.}},
\bauthor{\bsnm{Santamato}, \binits{E.}},
\bauthor{\bsnm{Nagali}, \binits{E.}},
\bauthor{\bsnm{Sciarrino}, \binits{F.}}:
\batitle{Spin-to-orbital optical angular momentum conversion in liquid crystal
  “q-plates”: Classical and quantum applications}.
\bjtitle{Molecular Crystals and Liquid Crystals}
\bvolume{561}(\bissue{1}),
\bfpage{48}--\blpage{56}
(\byear{2012})
\end{barticle}
\endbibitem

\bibitem[\protect\citeauthoryear{Kobashi et~al.}{2017}]{kobashi2017broadband}
\begin{barticle}
\bauthor{\bsnm{Kobashi}, \binits{J.}},
\bauthor{\bsnm{Yoshida}, \binits{H.}},
\bauthor{\bsnm{Ozaki}, \binits{M.}}:
\batitle{Broadband optical vortex generation from patterned cholesteric liquid
  crystals}.
\bjtitle{Molecular Crystals and Liquid Crystals}
\bvolume{646}(\bissue{1}),
\bfpage{116}--\blpage{124}
(\byear{2017})
\end{barticle}
\endbibitem

\bibitem[\protect\citeauthoryear{Ostrovsky
  et~al.}{2013}]{ostrovsky2013generation}
\begin{barticle}
\bauthor{\bsnm{Ostrovsky}, \binits{A.S.}},
\bauthor{\bsnm{Rickenstorff-Parrao}, \binits{C.}},
\bauthor{\bsnm{Arriz{\'o}n}, \binits{V.}}:
\batitle{Generation of the “perfect” optical vortex using a liquid-crystal
  spatial light modulator}.
\bjtitle{Optics letters}
\bvolume{38}(\bissue{4}),
\bfpage{534}--\blpage{536}
(\byear{2013})
\end{barticle}
\endbibitem

\bibitem[\protect\citeauthoryear{Mirhosseini
  et~al.}{2013}]{mirhosseini2013rapid}
\begin{barticle}
\bauthor{\bsnm{Mirhosseini}, \binits{M.}},
\bauthor{\bsnm{Magana-Loaiza}, \binits{O.S.}},
\bauthor{\bsnm{Chen}, \binits{C.}},
\bauthor{\bsnm{Rodenburg}, \binits{B.}},
\bauthor{\bsnm{Malik}, \binits{M.}},
\bauthor{\bsnm{Boyd}, \binits{R.W.}}:
\batitle{Rapid generation of light beams carrying orbital angular momentum}.
\bjtitle{Optics express}
\bvolume{21}(\bissue{25}),
\bfpage{30196}--\blpage{30203}
(\byear{2013})
\end{barticle}
\endbibitem

\bibitem[\protect\citeauthoryear{Hwang et~al.}{2024}]{hwang2024vortex}
\begin{barticle}
\bauthor{\bsnm{Hwang}, \binits{M.-S.}},
\bauthor{\bsnm{Kim}, \binits{H.-R.}},
\bauthor{\bsnm{Kim}, \binits{J.}},
\bauthor{\bsnm{Yang}, \binits{B.-J.}},
\bauthor{\bsnm{Kivshar}, \binits{Y.}},
\bauthor{\bsnm{Park}, \binits{H.-G.}}:
\batitle{Vortex nanolaser based on a photonic disclination cavity}.
\bjtitle{Nature Photonics}
\bvolume{18}(\bissue{3}),
\bfpage{286}--\blpage{293}
(\byear{2024})
\end{barticle}
\endbibitem

\bibitem[\protect\citeauthoryear{Bazhenov et~al.}{1990}]{bazhenov1990laser}
\begin{barticle}
\bauthor{\bsnm{Bazhenov}, \binits{V.Y.}},
\bauthor{\bsnm{Vasnetsov}, \binits{M.}},
\bauthor{\bsnm{Soskin}, \binits{M.}}:
\batitle{Laser beams with screw dislocations in their wavefronts}.
\bjtitle{Jetp Lett}
\bvolume{52}(\bissue{8}),
\bfpage{429}--\blpage{431}
(\byear{1990})
\end{barticle}
\endbibitem

\bibitem[\protect\citeauthoryear{Heckenberg
  et~al.}{1992}]{heckenberg1992generation}
\begin{barticle}
\bauthor{\bsnm{Heckenberg}, \binits{N.}},
\bauthor{\bsnm{McDuff}, \binits{R.}},
\bauthor{\bsnm{Smith}, \binits{C.}},
\bauthor{\bsnm{White}, \binits{A.}}:
\batitle{Generation of optical phase singularities by computer-generated
  holograms}.
\bjtitle{Optics letters}
\bvolume{17}(\bissue{3}),
\bfpage{221}--\blpage{223}
(\byear{1992})
\end{barticle}
\endbibitem

\bibitem[\protect\citeauthoryear{Basistiy et~al.}{1993}]{basistiy1993optics}
\begin{barticle}
\bauthor{\bsnm{Basistiy}, \binits{I.}},
\bauthor{\bsnm{Bazhenov}, \binits{V.Y.}},
\bauthor{\bsnm{Soskin}, \binits{M.}},
\bauthor{\bsnm{Vasnetsov}, \binits{M.V.}}:
\batitle{Optics of light beams with screw dislocations}.
\bjtitle{Optics communications}
\bvolume{103}(\bissue{5-6}),
\bfpage{422}--\blpage{428}
(\byear{1993})
\end{barticle}
\endbibitem

\bibitem[\protect\citeauthoryear{Berry and Dennis}{2001}]{berry2001knotted}
\begin{barticle}
\bauthor{\bsnm{Berry}, \binits{M.V.}},
\bauthor{\bsnm{Dennis}, \binits{M.R.}}:
\batitle{Knotted and linked phase singularities in monochromatic waves}.
\bjtitle{Proceedings of the Royal Society of London. Series A: Mathematical,
  Physical and Engineering Sciences}
\bvolume{457}(\bissue{2013}),
\bfpage{2251}--\blpage{2263}
(\byear{2001})
\end{barticle}
\endbibitem

\bibitem[\protect\citeauthoryear{Ricci et~al.}{2012}]{ricci2012instability}
\begin{barticle}
\bauthor{\bsnm{Ricci}, \binits{F.}},
\bauthor{\bsnm{L{\"o}ffler}, \binits{W.}},
\bauthor{\bsnm{Van~Exter}, \binits{M.}}:
\batitle{Instability of higher-order optical vortices analyzed with a
  multi-pinhole interferometer}.
\bjtitle{Optics express}
\bvolume{20}(\bissue{20}),
\bfpage{22961}--\blpage{22975}
(\byear{2012})
\end{barticle}
\endbibitem

\bibitem[\protect\citeauthoryear{Abramochkin and
  Volostnikov}{1991}]{abramochkin1991beam}
\begin{barticle}
\bauthor{\bsnm{Abramochkin}, \binits{E.}},
\bauthor{\bsnm{Volostnikov}, \binits{V.}}:
\batitle{Beam transformations and nontransformed beams}.
\bjtitle{Optics Communications}
\bvolume{83}(\bissue{1-2}),
\bfpage{123}--\blpage{135}
(\byear{1991})
\end{barticle}
\endbibitem

\bibitem[\protect\citeauthoryear{Vaity et~al.}{2013}]{vaity2013measuring}
\begin{barticle}
\bauthor{\bsnm{Vaity}, \binits{P.}},
\bauthor{\bsnm{Banerji}, \binits{J.}},
\bauthor{\bsnm{Singh}, \binits{R.}}:
\batitle{Measuring the topological charge of an optical vortex by using a
  tilted convex lens}.
\bjtitle{Physics letters a}
\bvolume{377}(\bissue{15}),
\bfpage{1154}--\blpage{1156}
(\byear{2013})
\end{barticle}
\endbibitem

\bibitem[\protect\citeauthoryear{Denisenko
  et~al.}{2009}]{denisenko2009determination}
\begin{barticle}
\bauthor{\bsnm{Denisenko}, \binits{V.}},
\bauthor{\bsnm{Shvedov}, \binits{V.}},
\bauthor{\bsnm{Desyatnikov}, \binits{A.S.}},
\bauthor{\bsnm{Neshev}, \binits{D.N.}},
\bauthor{\bsnm{Krolikowski}, \binits{W.}},
\bauthor{\bsnm{Volyar}, \binits{A.}},
\bauthor{\bsnm{Soskin}, \binits{M.}},
\bauthor{\bsnm{Kivshar}, \binits{Y.S.}}:
\batitle{Determination of topological charges of polychromatic optical
  vortices}.
\bjtitle{Optics express}
\bvolume{17}(\bissue{26}),
\bfpage{23374}--\blpage{23379}
(\byear{2009})
\end{barticle}
\endbibitem

\bibitem[\protect\citeauthoryear{Forbes et~al.}{2016}]{forbes2016creation}
\begin{barticle}
\bauthor{\bsnm{Forbes}, \binits{A.}},
\bauthor{\bsnm{Dudley}, \binits{A.}},
\bauthor{\bsnm{McLaren}, \binits{M.}}:
\batitle{Creation and detection of optical modes with spatial light
  modulators}.
\bjtitle{Advances in optics and photonics}
\bvolume{8}(\bissue{2}),
\bfpage{200}--\blpage{227}
(\byear{2016})
\end{barticle}
\endbibitem

\bibitem[\protect\citeauthoryear{Kotlyar et~al.}{1998}]{kotlyar1998light}
\begin{barticle}
\bauthor{\bsnm{Kotlyar}, \binits{V.}},
\bauthor{\bsnm{Khonina}, \binits{S.}},
\bauthor{\bsnm{Soifer}, \binits{V.}}:
\batitle{Light field decomposition in angular harmonics by means of diffractive
  optics}.
\bjtitle{Journal of modern optics}
\bvolume{45}(\bissue{7}),
\bfpage{1495}--\blpage{1506}
(\byear{1998})
\end{barticle}
\endbibitem

\bibitem[\protect\citeauthoryear{Dai et~al.}{2015}]{dai2015measuring}
\begin{barticle}
\bauthor{\bsnm{Dai}, \binits{K.}},
\bauthor{\bsnm{Gao}, \binits{C.}},
\bauthor{\bsnm{Zhong}, \binits{L.}},
\bauthor{\bsnm{Na}, \binits{Q.}},
\bauthor{\bsnm{Wang}, \binits{Q.}}:
\batitle{Measuring oam states of light beams with gradually-changing-period
  gratings}.
\bjtitle{Optics letters}
\bvolume{40}(\bissue{4}),
\bfpage{562}--\blpage{565}
(\byear{2015})
\end{barticle}
\endbibitem

\bibitem[\protect\citeauthoryear{Guo et~al.}{2009}]{guo2009measuring}
\begin{botherref}
\oauthor{\bsnm{Guo}, \binits{C.-S.}},
\oauthor{\bsnm{Yue}, \binits{S.-J.}},
\oauthor{\bsnm{Wei}, \binits{G.-X.}}:
Measuring the orbital angular momentum of optical vortices using a multipinhole
  plate.
Applied Physics Letters
\textbf{94}(23)
(2009)
\end{botherref}
\endbibitem

\bibitem[\protect\citeauthoryear{Anderson et~al.}{2012}]{anderson2012measuring}
\begin{barticle}
\bauthor{\bsnm{Anderson}, \binits{M.E.}},
\bauthor{\bsnm{Bigman}, \binits{H.}},
\bauthor{\bsnm{Araujo}, \binits{L.E.}},
\bauthor{\bsnm{Chaloupka}, \binits{J.L.}}:
\batitle{Measuring the topological charge of ultrabroadband, optical-vortex
  beams with a triangular aperture}.
\bjtitle{Journal of the Optical Society of America B}
\bvolume{29}(\bissue{8}),
\bfpage{1968}--\blpage{1976}
(\byear{2012})
\end{barticle}
\endbibitem

\bibitem[\protect\citeauthoryear{Leach et~al.}{2002}]{leach2002measuring}
\begin{barticle}
\bauthor{\bsnm{Leach}, \binits{J.}},
\bauthor{\bsnm{Padgett}, \binits{M.J.}},
\bauthor{\bsnm{Barnett}, \binits{S.M.}},
\bauthor{\bsnm{Franke-Arnold}, \binits{S.}},
\bauthor{\bsnm{Courtial}, \binits{J.}}:
\batitle{Measuring the orbital angular momentum of a single photon}.
\bjtitle{Physical review letters}
\bvolume{88}(\bissue{25}),
\bfpage{257901}
(\byear{2002})
\end{barticle}
\endbibitem

\bibitem[\protect\citeauthoryear{Fr{\k{a}}czek et~al.}{2006}]{frkaczek2006new}
\begin{barticle}
\bauthor{\bsnm{Fr{\k{a}}czek}, \binits{E.}},
\bauthor{\bsnm{Fr{\k{a}}czek}, \binits{W.}},
\bauthor{\bsnm{Masajada}, \binits{J.}}:
\batitle{The new method of topological charge determination of optical vortices
  in the interference field of the optical vortex interferometer}.
\bjtitle{Optik}
\bvolume{117}(\bissue{9}),
\bfpage{423}--\blpage{425}
(\byear{2006})
\end{barticle}
\endbibitem

\bibitem[\protect\citeauthoryear{Mikulich et~al.}{2016}]{mikulich2016influence}
\begin{barticle}
\bauthor{\bsnm{Mikulich}, \binits{V.}},
\bauthor{\bsnm{Murawski}, \binits{A.A.}},
\bauthor{\bsnm{Muravsky}, \binits{A.A.}},
\bauthor{\bsnm{Agabekov}, \binits{V.}}:
\batitle{Influence of methyl substituents on azo-dye photoalignment in thin
  films}.
\bjtitle{Journal of Applied Spectroscopy}
\bvolume{83},
\bfpage{115}--\blpage{120}
(\byear{2016})
\end{barticle}
\endbibitem

\bibitem[\protect\citeauthoryear{Muravsky
  et~al.}{2020}]{muravsky2020photoinduced}
\begin{barticle}
\bauthor{\bsnm{Muravsky}, \binits{A.}},
\bauthor{\bsnm{Murauski}, \binits{A.}},
\bauthor{\bsnm{Kukhta}, \binits{I.}}:
\batitle{Photoinduced hole dipoles’ mechanism of liquid crystal
  photoalignment}.
\bjtitle{Applied Optics}
\bvolume{59}(\bissue{17}),
\bfpage{5102}--\blpage{5107}
(\byear{2020})
\end{barticle}
\endbibitem

\bibitem[\protect\citeauthoryear{Kazak et~al.}{2010}]{kazak2010controlling}
\begin{barticle}
\bauthor{\bsnm{Kazak}, \binits{A.}},
\bauthor{\bsnm{Tolstik}, \binits{A.}},
\bauthor{\bsnm{Mel'nikova}, \binits{E.}}:
\batitle{Controlling light fields by means of liquid-crystal diffraction
  elements}.
\bjtitle{Journal of Optical Technology}
\bvolume{77}(\bissue{7}),
\bfpage{461}--\blpage{462}
(\byear{2010})
\end{barticle}
\endbibitem

\bibitem[\protect\citeauthoryear{Chigrinov
  et~al.}{2020}]{chigrinov2020photoaligning}
\begin{barticle}
\bauthor{\bsnm{Chigrinov}, \binits{V.}},
\bauthor{\bsnm{Sun}, \binits{J.}},
\bauthor{\bsnm{Wang}, \binits{X.}}:
\batitle{Photoaligning and photopatterning: New lc technology}.
\bjtitle{Crystals}
\bvolume{10}(\bissue{4}),
\bfpage{323}
(\byear{2020})
\end{barticle}
\endbibitem

\bibitem[\protect\citeauthoryear{Wang et~al.}{2013}]{wang2013switchable}
\begin{barticle}
\bauthor{\bsnm{Wang}, \binits{X.-Q.}},
\bauthor{\bsnm{Srivastava}, \binits{A.K.}},
\bauthor{\bsnm{Chigrinov}, \binits{V.G.}},
\bauthor{\bsnm{Kwok}, \binits{H.-S.}}:
\batitle{Switchable fresnel lens based on micropatterned alignment}.
\bjtitle{Optics Letters}
\bvolume{38}(\bissue{11}),
\bfpage{1775}--\blpage{1777}
(\byear{2013})
\end{barticle}
\endbibitem

\bibitem[\protect\citeauthoryear{Melnikova
  et~al.}{2022}]{melnikova2022polarization}
\begin{barticle}
\bauthor{\bsnm{Melnikova}, \binits{E.}},
\bauthor{\bsnm{Stashkevich}, \binits{I.}},
\bauthor{\bsnm{Rushnova}, \binits{I.}},
\bauthor{\bsnm{Tolstik}, \binits{A.}},
\bauthor{\bsnm{Timofeev}, \binits{S.}}:
\batitle{Polarization properties of the electrically controlled twist-planar
  liquid crystal di. raction structure}.
\bjtitle{Nonlinear Phenomena in Complex Systems}
\bvolume{25}(\bissue{3}),
\bfpage{229}--\blpage{244}
(\byear{2022})
\end{barticle}
\endbibitem

\bibitem[\protect\citeauthoryear{W{\k{e}}g{\l}owski
  et~al.}{2017}]{wkeglowski2017electro}
\begin{barticle}
\bauthor{\bsnm{W{\k{e}}g{\l}owski}, \binits{R.}},
\bauthor{\bsnm{Kozanecka-Szmigiel}, \binits{A.}},
\bauthor{\bsnm{Piecek}, \binits{W.}},
\bauthor{\bsnm{Konieczkowska}, \binits{J.}},
\bauthor{\bsnm{Schab-Balcerzak}, \binits{E.}}:
\batitle{Electro-optically tunable diffraction grating with photoaligned liquid
  crystals}.
\bjtitle{Optics Communications}
\bvolume{400},
\bfpage{144}--\blpage{149}
(\byear{2017})
\end{barticle}
\endbibitem

\bibitem[\protect\citeauthoryear{Zhou et~al.}{2015}]{zhou2015optical}
\begin{botherref}
\oauthor{\bsnm{Zhou}, \binits{H.}},
\oauthor{\bsnm{Choate}, \binits{E.P.}},
\oauthor{\bsnm{Wang}, \binits{H.}}:
Optical fredericks transition in a nematic liquid crystal layer.
Liquid Crystalline Polymers: Volume 2--Processing and Applications,
265--295
(2015)
\end{botherref}
\endbibitem

\bibitem[\protect\citeauthoryear{Lucchetti et~al.}{2014}]{lucchetti2014light}
\begin{botherref}
\oauthor{\bsnm{Lucchetti}, \binits{L.}},
\oauthor{\bsnm{Catani}, \binits{L.}},
\oauthor{\bsnm{Simoni}, \binits{F.}}:
Light-controlled electric freedericksz threshold in dye doped liquid crystals.
Journal of Applied Physics
\textbf{115}(20)
(2014)
\end{botherref}
\endbibitem

\bibitem[\protect\citeauthoryear{Mauguin}{1911}]{mauguin1911cristaux}
\begin{barticle}
\bauthor{\bsnm{Mauguin}, \binits{C.}}:
\batitle{Sur les cristaux liquides de m. lehmann}.
\bjtitle{Bulletin de Min{\'e}ralogie}
\bvolume{34}(\bissue{3}),
\bfpage{71}--\blpage{117}
(\byear{1911})
\end{barticle}
\endbibitem

\bibitem[\protect\citeauthoryear{Mel’nikova
  et~al.}{2024}]{mel2024electrically}
\begin{botherref}
\oauthor{\bsnm{Mel’nikova}, \binits{E.}},
\oauthor{\bsnm{Panteleeva}, \binits{E.}},
\oauthor{\bsnm{Gorbach}, \binits{D.}},
\oauthor{\bsnm{Tolstik}, \binits{A.}},
\oauthor{\bsnm{Rushnova}, \binits{I.}},
\oauthor{\bsnm{Kabanova}, \binits{O.}}:
Electrically controlled liquid crystal twist-planar fresnel lens.
Journal of Applied Spectroscopy,
1--6
(2024)
\end{botherref}
\endbibitem

\bibitem[\protect\citeauthoryear{Soskin et~al.}{1997}]{soskin1997topological}
\begin{barticle}
\bauthor{\bsnm{Soskin}, \binits{M.}},
\bauthor{\bsnm{Gorshkov}, \binits{V.}},
\bauthor{\bsnm{Vasnetsov}, \binits{M.}},
\bauthor{\bsnm{Malos}, \binits{J.}},
\bauthor{\bsnm{Heckenberg}, \binits{N.}}:
\batitle{Topological charge and angular momentum of light beams carrying
  optical vortices}.
\bjtitle{Physical Review A}
\bvolume{56}(\bissue{5}),
\bfpage{4064}
(\byear{1997})
\end{barticle}
\endbibitem

\bibitem[\protect\citeauthoryear{Ma et~al.}{2013}]{ma2013study}
\begin{barticle}
\bauthor{\bsnm{Ma}, \binits{H.}},
\bauthor{\bsnm{Hu}, \binits{H.}},
\bauthor{\bsnm{Xie}, \binits{W.}},
\bauthor{\bsnm{Xu}, \binits{X.}}:
\batitle{Study on the generation of a vortex laser beam by using phase-only
  liquid crystal spatial light modulator}.
\bjtitle{Optical Engineering}
\bvolume{52}(\bissue{9}),
\bfpage{091721}--\blpage{091721}
(\byear{2013})
\end{barticle}
\endbibitem

\bibitem[\protect\citeauthoryear{Szatkowski
  et~al.}{2020}]{szatkowski2020generation}
\begin{barticle}
\bauthor{\bsnm{Szatkowski}, \binits{M.}},
\bauthor{\bsnm{Masajada}, \binits{J.}},
\bauthor{\bsnm{Augustyniak}, \binits{I.}},
\bauthor{\bsnm{Nowacka}, \binits{K.}}:
\batitle{Generation of composite vortex beams by independent spatial light
  modulator pixel addressing}.
\bjtitle{Optics Communications}
\bvolume{463},
\bfpage{125341}
(\byear{2020})
\end{barticle}
\endbibitem

\bibitem[\protect\citeauthoryear{Nersisyan
  et~al.}{2009}]{nersisyan2009fabrication}
\begin{barticle}
\bauthor{\bsnm{Nersisyan}, \binits{S.}},
\bauthor{\bsnm{Tabiryan}, \binits{N.}},
\bauthor{\bsnm{Steeves}, \binits{D.M.}},
\bauthor{\bsnm{Kimball}, \binits{B.R.}}:
\batitle{Fabrication of liquid crystal polymer axial waveplates for uv-ir
  wavelengths}.
\bjtitle{Optics express}
\bvolume{17}(\bissue{14}),
\bfpage{11926}--\blpage{11934}
(\byear{2009})
\end{barticle}
\endbibitem

\bibitem[\protect\citeauthoryear{Slussarenko
  et~al.}{2011}]{slussarenko2011tunable}
\begin{barticle}
\bauthor{\bsnm{Slussarenko}, \binits{S.}},
\bauthor{\bsnm{Murauski}, \binits{A.}},
\bauthor{\bsnm{Du}, \binits{T.}},
\bauthor{\bsnm{Chigrinov}, \binits{V.}},
\bauthor{\bsnm{Marrucci}, \binits{L.}},
\bauthor{\bsnm{Santamato}, \binits{E.}}:
\batitle{Tunable liquid crystal q-plates with arbitrary topological charge}.
\bjtitle{Optics express}
\bvolume{19}(\bissue{5}),
\bfpage{4085}--\blpage{4090}
(\byear{2011})
\end{barticle}
\endbibitem

\bibitem[\protect\citeauthoryear{Courtial
  et~al.}{1998}]{courtial1998measurement}
\begin{barticle}
\bauthor{\bsnm{Courtial}, \binits{J.}},
\bauthor{\bsnm{Dholakia}, \binits{K.}},
\bauthor{\bsnm{Robertson}, \binits{D.}},
\bauthor{\bsnm{Allen}, \binits{L.}},
\bauthor{\bsnm{Padgett}, \binits{M.}}:
\batitle{Measurement of the rotational frequency shift imparted to a rotating
  light beam possessing orbital angular momentum}.
\bjtitle{Physical review letters}
\bvolume{80}(\bissue{15}),
\bfpage{3217}
(\byear{1998})
\end{barticle}
\endbibitem

\bibitem[\protect\citeauthoryear{Harris et~al.}{1994}]{harris1994laser}
\begin{barticle}
\bauthor{\bsnm{Harris}, \binits{M.}},
\bauthor{\bsnm{Hill}, \binits{C.}},
\bauthor{\bsnm{Tapster}, \binits{P.}},
\bauthor{\bsnm{Vaughan}, \binits{J.}}:
\batitle{Laser modes with helical wave fronts}.
\bjtitle{Physical Review A}
\bvolume{49}(\bissue{4}),
\bfpage{3119}
(\byear{1994})
\end{barticle}
\endbibitem

\bibitem[\protect\citeauthoryear{Cui et~al.}{2019}]{cui2019determining}
\begin{barticle}
\bauthor{\bsnm{Cui}, \binits{S.}},
\bauthor{\bsnm{Xu}, \binits{B.}},
\bauthor{\bsnm{Luo}, \binits{S.}},
\bauthor{\bsnm{Xu}, \binits{H.}},
\bauthor{\bsnm{Cai}, \binits{Z.}},
\bauthor{\bsnm{Luo}, \binits{Z.}},
\bauthor{\bsnm{Pu}, \binits{J.}},
\bauthor{\bsnm{Ch{\'a}vez-Cerda}, \binits{S.}}:
\batitle{Determining topological charge based on an improved fizeau
  interferometer}.
\bjtitle{Optics express}
\bvolume{27}(\bissue{9}),
\bfpage{12774}--\blpage{12779}
(\byear{2019})
\end{barticle}
\endbibitem

\bibitem[\protect\citeauthoryear{Melo et~al.}{2018}]{melo2018direct}
\begin{barticle}
\bauthor{\bsnm{Melo}, \binits{L.A.}},
\bauthor{\bsnm{Jesus-Silva}, \binits{A.J.}},
\bauthor{\bsnm{Ch{\'a}vez-Cerda}, \binits{S.}},
\bauthor{\bsnm{Ribeiro}, \binits{P.H.S.}},
\bauthor{\bsnm{Soares}, \binits{W.C.}}:
\batitle{Direct measurement of the topological charge in elliptical beams using
  diffraction by a triangular aperture}.
\bjtitle{Scientific Reports}
\bvolume{8}(\bissue{1}),
\bfpage{6370}
(\byear{2018})
\end{barticle}
\endbibitem

\bibitem[\protect\citeauthoryear{Han and Zhao}{2011}]{han2011measuring}
\begin{barticle}
\bauthor{\bsnm{Han}, \binits{Y.}},
\bauthor{\bsnm{Zhao}, \binits{G.}}:
\batitle{Measuring the topological charge of optical vortices with an axicon}.
\bjtitle{Optics letters}
\bvolume{36}(\bissue{11}),
\bfpage{2017}--\blpage{2019}
(\byear{2011})
\end{barticle}
\endbibitem

\bibitem[\protect\citeauthoryear{Izdebskaya
  et~al.}{2012}]{izdebskaya2012observation}
\begin{barticle}
\bauthor{\bsnm{Izdebskaya}, \binits{Y.V.}},
\bauthor{\bsnm{Rebling}, \binits{J.}},
\bauthor{\bsnm{Desyatnikov}, \binits{A.S.}},
\bauthor{\bsnm{Kivshar}, \binits{Y.S.}}:
\batitle{Observation of vector solitons with hidden vorticity}.
\bjtitle{Optics Letters}
\bvolume{37}(\bissue{5}),
\bfpage{767}--\blpage{769}
(\byear{2012})
\end{barticle}
\endbibitem

\bibitem[\protect\citeauthoryear{Mesquita
  et~al.}{2011}]{mesquita2011engineering}
\begin{barticle}
\bauthor{\bsnm{Mesquita}, \binits{P.H.}},
\bauthor{\bsnm{Jesus-Silva}, \binits{A.J.}},
\bauthor{\bsnm{Fonseca}, \binits{E.J.}},
\bauthor{\bsnm{Hickmann}, \binits{J.M.}}:
\batitle{Engineering a square truncated lattice with light's orbital angular
  momentum}.
\bjtitle{Optics express}
\bvolume{19}(\bissue{21}),
\bfpage{20616}--\blpage{20621}
(\byear{2011})
\end{barticle}
\endbibitem

\bibitem[\protect\citeauthoryear{Muravsky et~al.}{2021}]{muravsky2021high}
\begin{barticle}
\bauthor{\bsnm{Muravsky}, \binits{A.A.}},
\bauthor{\bsnm{Murauski}, \binits{A.A.}},
\bauthor{\bsnm{Kukhta}, \binits{I.N.}},
\bauthor{\bsnm{Yakovleva}, \binits{A.S.}}:
\batitle{High anchoring photoalignment material based on new photo-induced hole
  dipoles' mechanism}.
\bjtitle{Journal of the Society for Information Display}
\bvolume{29}(\bissue{11}),
\bfpage{833}--\blpage{839}
(\byear{2021})
\end{barticle}
\endbibitem

\bibitem[\protect\citeauthoryear{Melnikova et~al.}{2022}]{melnikova2022liquid}
\begin{barticle}
\bauthor{\bsnm{Melnikova}, \binits{E.}},
\bauthor{\bsnm{Gorbach}, \binits{D.}},
\bauthor{\bsnm{Slussarenko~Sr}, \binits{S.}},
\bauthor{\bsnm{Muravsky}, \binits{A.}},
\bauthor{\bsnm{Tolstik}, \binits{A.}},
\bauthor{\bsnm{Slussarenko~Jr}, \binits{S.}}:
\batitle{Liquid-crystal q-plates with a phase core to generation vortex beams
  with controllable number of singularities}.
\bjtitle{Optics Communications}
\bvolume{522},
\bfpage{128661}
(\byear{2022})
\end{barticle}
\endbibitem

\bibitem[\protect\citeauthoryear{Melnikova}{2022}]{melnikova2022electrically}
\begin{barticle}
\bauthor{\bsnm{Melnikova}, \binits{E.A.}}:
\batitle{Electrically controlled microstructured liquid-crystal twist elements
  for phase conversion of light fields}.
\bjtitle{Journal of Optical Technology}
\bvolume{89}(\bissue{3}),
\bfpage{169}--\blpage{175}
(\byear{2022})
\end{barticle}
\endbibitem

\bibitem[\protect\citeauthoryear{Melnikova et~al.}{2024}]{melnikova2024tunable}
\begin{botherref}
\oauthor{\bsnm{Melnikova}, \binits{E.}},
\oauthor{\bsnm{Pantsialeyeva}, \binits{Y.}},
\oauthor{\bsnm{Gorbach}, \binits{D.}},
\oauthor{\bsnm{Tolstik}, \binits{A.}},
\oauthor{\bsnm{Karabchevsky}, \binits{A.}}:
Tunable liquid crystal twisted-planar fresnel lens for vortex topology
  determination.
Laser \& Photonics Reviews,
2401006
(2024)
\end{botherref}
\endbibitem

\bibitem[\protect\citeauthoryear{Plick and Krenn}{2015}]{plick2015physical}
\begin{barticle}
\bauthor{\bsnm{Plick}, \binits{W.N.}},
\bauthor{\bsnm{Krenn}, \binits{M.}}:
\batitle{Physical meaning of the radial index of laguerre-gauss beams}.
\bjtitle{Physical Review A}
\bvolume{92}(\bissue{6}),
\bfpage{063841}
(\byear{2015})
\end{barticle}
\endbibitem

\end{thebibliography}

	\appendix
	\section*{Supplementary Information}
	\renewcommand{\thefigure}{S\arabic{figure}}
	\setcounter{figure}{0}
	
	\renewcommand{\thesubsection}{\arabic{subsection}}
	\makeatletter
	\def\@seccntformat#1{\@ifundefined{#1@cntformat}
		{\csname the#1\endcsname\quad}
		{\csname #1@cntformat\endcsname}}
	\newcommand\section@cntformat{}   
	\makeatother
	\subsection{Fabrication of the Electrically Controlled Twist-Planar Oriented Liquid Crystal Fresnel Lens}
	
	The method of manufacturing the NLC Fresnel lens, the structural scheme of which is shown in Figure \ref{fig:Schematic diagram of the element}a, b, consists of several successive stages: preparation of glass substrates, deposition of thin films of AtA-2 azo dye on the substrates, photoorientation of AtA-2 azo dye, cell assembly, filling of the cell with nematic liquid crystal.
	
	\textit{Substrates Preparation}: the corpus of the controlled NLC Fresnel lens was manufactured using 1.1 mm thick glass substrates with a continuous transparent electrically conductive indium tin oxide (ITO) coating with a resistance of 100 Ohm square$^{-1}$ on the inner surface (INTEGRAL OJSC, Minsk, Belarus). A diamond roller glass cutter (Bohle AG) was used to prepare the substrates of the required size (20x30 mm) and shape.
	
	The pre-cleaning of the substrate surfaces took place in an ultrasonic bath: once in a solution of a surfactant with distilled water at a temperature of T = 45°C for 15 minutes; then two times in distilled water at a temperature of T = 45°C for 15 minutes. The next step was to clean the glass substrates with a cotton textile soaked in isopropyl alcohol. The substrates are then dried with a second clean, dry cotton textile. At this stage of cleaning, dust particles and lint from the cleaning cloth are allowed on the glass (dirt and stains should not be present). To remove foreign particles from the substrate surface, they are blown with a jet of compressed air using a compressor (Remeza CJSC, Minsk, Belarus).
	
	To check the degree of purification, two substrates are pressed tightly against each other with their work surfaces inside, and interference fringes are observed in the light of a table lamp. If there are few of them (2-4), then it is assumed that there are no foreign particles and dirt on the work surfaces. Otherwise, it is necessary to repeat the process of cleaning the working surfaces of glass substrates.
	
	Organic pollutants were removed using a UV cleaning system (Photo Surface Processor PL16-110D, SEN LIGHTS Corp., JP). After application of the system, it is necessary to blow the substrates again with compressed air to remove traces of organic pollutants.
	
	\textit{Deposition of AtA-2 Azo Dye Thin Films to Substrates}: to deposite thin (about 30 nm) films of photoorienting material (AtA-2 azo dye synthesized in the Laboratory of Materials and Technologies of LC Devices of the Institute of Chemistry of New Materials of the National Academy of Sciences of Belarus) from liquid solutions to the inner surfaces of purified glass substrates, the method of creating a meniscus with a rod (Mayer-Rod Coating) was used. AtA-2 azo dye has high sensitivity in the blue region of the spectrum and is characterized by high adhesion energy to LC molecules \cite{mikulich2016influence, muravsky2021high}, implements a new mechanism for the orientation of a liquid crystal of "photoinduced hole dipoles" \cite{muravsky2020photoinduced}. After deposition, the azo dye must be dried on a hot plate at 140°C for 5 minutes to evaporate the solvent.
	
	\textit{Photoorientation of the AtA-2 Azo Dye}: azo dye films acquire orienting properties when exposed to linearly polarized light. It is important to note that the direction of the induced orientation of the azo dye molecules is perpendicular to the direction of polarization of the activating radiation. The orientation direction in the AtA-2 azo dye thin film can be easily changed further by additional irradiation with activating radiation with an alternate polarization direction. 
	
	The activation of ATA-2 azo dye on the inner surfaces of glass substrates was carried out in two stages \cite{melnikova2022liquid, melnikova2022electrically}. At the first stage, both substrates were uniformly exposed with LED radiation with a wavelength $\lambda$ = 450 nm and linear polarization (the power density was I = 15 mW cm$^{-2}$). At the second stage, a thin film of the photoorientant was re-exposed on only one substrate using an amplitude glass mask and linearly polarized radiation with an orthogonal direction of the radiation polarization vector relative to its direction at the first stage and a power density of I = 45 mW cm$^{-2}$. The transmission profile of the amplitude glass mask for photoorientation of AtA-2 azo dye corresponded to a Fresnel lens with a radius of the first zone $r_1$ = 333 $\mu$m. The mask was made by lithographic technology on OJSC Integral (Minsk, Belarus). 
	
	\textit{Assembling the Lens Cell}: the body of the NLC Fresnel lens was assembled in a laminar flow cabinet equipped with an air flow circulation and purification system in order to avoid repeated contamination of the working surfaces of the glass substrates. The thickness of the homogeneous air gap in the experimental sample was determined by fiber polymer spacers with a diameter of 20 $\mu$m located along two opposite edges of the substrates. The glass substrates were glued along the perimeter with UV glue (Norland NOA81, Edmund Optics) from two opposite sides, leaving space on the substrates for voltage supply and cell control.
	
	The formed air gap between the glued substrates makes it possible to fill the element with a nematic liquid crystal by the capillary method under the condition of its isotropic phase at a temperature of 65°C in the drying cabinet (laboratory furnaces SNOL 58/350, AB UMEGA-GROUP Lithuania). 
	
	\subsection{Experimental Section}
	
	\textit{Electrical Control of the Properties of the NLC Fresnel Lens}: to obtain the electro-optical response of the NLC layer, which provides effective electrical control of diffraction properties, an alternating voltage of rectangular pulse shape (meander) with a frequency of 1 kHz was applied to the electrodes of the NLC lens using an installation Arbitrary Waveform Generator B-332 (LANFOR Company, St. Petersburg, Russia).
	
	\textit{Polarization Microscopy} (Figure \ref{fig:Schematic diagram of the element}c): the study of the formed anisotropic structure in the NLC layer was carried out on video complex based on polarizing microscope (Optoelectronic Systems OJSC, Minsk, Belarus) with digital camera, resolution of 2560x1920 pixels.
	
	\textit{Experimental Systems for Phase Topology Measurements} (Figure \ref{fig:Scheme}b): in the work, radiation with a wavelength of $\lambda$=632.8 nm was used, generated by He-Ne laser LGK 7665 P18 (LASOS Lasertechnik GmbH). The size matching of the laser radiation and the Fresnel LC lens (about 4 mm) was implemented using a collimator consisting of a micro lens (20x) and a focusing lens (f=20cm).
	
	The formation of a phase singular beam with a topological charge $\ell$ from $\pm1$ to $\pm8$ was realized using a Spiral Plate (SP) VPP-m633 (RPC Photonics, Inc.).
	
\begin{figure}[h]
		\centering
		\includegraphics[width=0.7\linewidth]{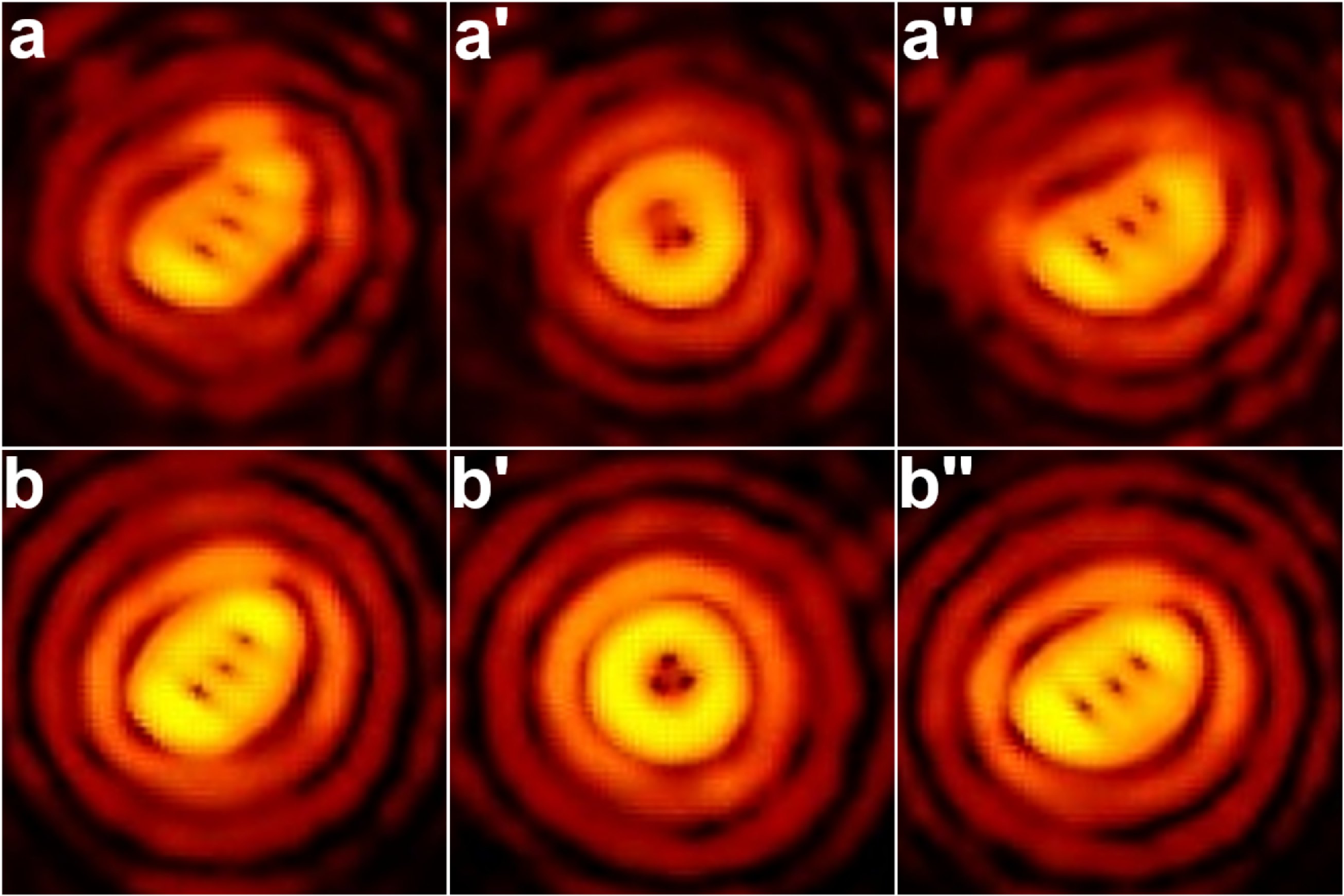}
		\caption{$\mid$ \textbf{Experimental characterization of the phase topology of OVs via NLC Fresnel lens.} The distribution profile of the light field of a phase singular beam with topological charge $\ell=+3$ in the focal plane of the Fresnel lens: \textbf{a$^\prime$}, \textbf{b$^\prime$}, under normal incidence of an optical vortex on the NLC lens ($\theta$ = 0 degrees) and when the radiation incidence at angles of $\theta=-10$ degrees (\textbf{a}, \textbf{b}) and $\theta=+10$ degrees (\textbf{a$^{\prime\prime}$}, \textbf{b$^{\prime\prime}$}) with control voltage \textit{U} = 0 V (a-a$^{\prime\prime}$) and \textit{U} = 3 V (b-b$^{\prime\prime}$).}
		\label{fig:Supl1}
	\end{figure}

	Intensity distribution profiles (Figures \ref{fig:Scheme}, \ref{fig:All}, \ref{fig:Angle2}a, \ref{fig:focal plane}, \ref{fig:focal plane2}a$^{\prime}$-a$^{\prime\prime\prime}$, b$^{\prime}$-b$^{\prime\prime\prime}$, c$^{\prime}$-c$^{\prime\prime\prime}$, \ref{fig:Supl1}, \ref{fig:Supl2}) were recorded using a CCD camera FLIRCHAMELEON CMLN-13S2M-CS (FLIR Integrated Imaging Solutions, Canada). 
	
	The experimental results were confirmed using a focusing spherical lens with a focal length of f=17cm (Figure \ref{fig:focal plane2}a$^{\prime\prime\prime}$- c$^{\prime\prime\prime}$, \ref{fig:Supl2}).
	
	Figure \ref{fig:Supl1} shows the experimental results of the astigmatic transformation of an optical vortex with a topological charge $\ell=+3$ at the external control voltage (\textit{U}) of 0 V and 3 V ("topological charge detection" mode) to the NLC Fresnel lens. It can be seen from the figure that in the absence of rotation of the lens in the Fourier plane, the optical vortex splits into three phase singular beams with topological charges $\mid\ell\mid$ = 1. The rotation of the NLC Fresnel lens around the Z axis leads to the alignment of intensity zeros on a straight line. Applying the optimal voltage (3 V) to the NLC element results in an increase in intensity.
	
	\subsection{Analysis in the Mach-Zehnder Interferometer}
	
\begin{figure}[H]
		\centering
		\includegraphics[width=\linewidth]{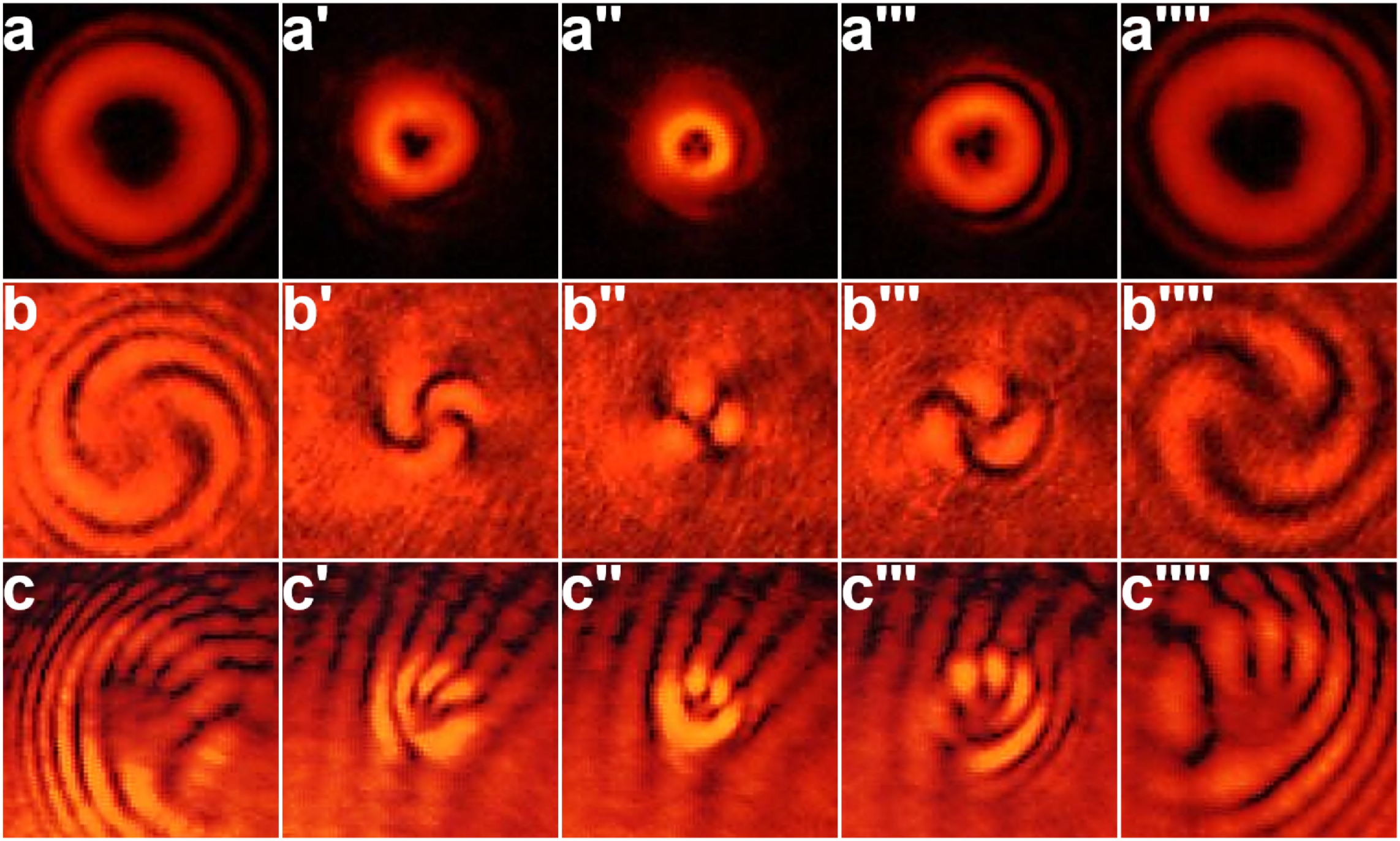}
		\caption{$\mid$ \textbf{Analysis in the Mach-Zehnder interferometer.} Light field distribution profile: 2 cm before (\textbf{a-c}), 0.6 cm before (\textbf{a$^\prime$-c$^\prime$}), in (\textbf{a$^{\prime\prime}$-c$^{\prime\prime}$}), 0.6 cm after (\textbf{a$^{\prime\prime\prime}$-c$^{\prime\prime\prime}$}) and 2 cm after (\textbf{a$^{\prime\prime\prime\prime}$-c$^{\prime\prime\prime\prime}$}) the focal plane of the spherical lens. \textbf{a-a$^{\prime\prime\prime\prime}$}, photographs of optical vortex with a topological charge $\ell$ = +3 when passing through a spherical lens. The photographs of the interference pattern of an optical vortex with a spherical wave (b-b$^{\prime\prime\prime\prime}$) and a plane wave (c-c$^{\prime\prime\prime\prime}$). }
		\label{fig:Supl2}
	\end{figure}

	A study of the phase topology of the vortex using a Mach-Zehnder interferometer (the method of coherent addition of the studied beam with a spherical  (Figure \ref{fig:Supl2}b-b$^{\prime\prime\prime\prime}$) or plane wave  (Figure \ref{fig:Supl2}c-c$^{\prime\prime\prime\prime}$)) showed a change in the sign of the topological charge of the phase singular beam focused by the lens to the opposite after passing through the focal plane of the lens \cite{melnikova2024tunable}. The instability of a phase singular beam in the Fourier plane of the spherical lens has also been experimentally shown (Figure \ref{fig:Supl2}a$^{\prime\prime}$-c$^{\prime\prime}$).
	
	\subsection{Numerical Simulations of Vortex Beam Propagation via the NLC Fresnel Lens}
	
\begin{figure}[H]
		\centering
		\includegraphics[width=\linewidth]{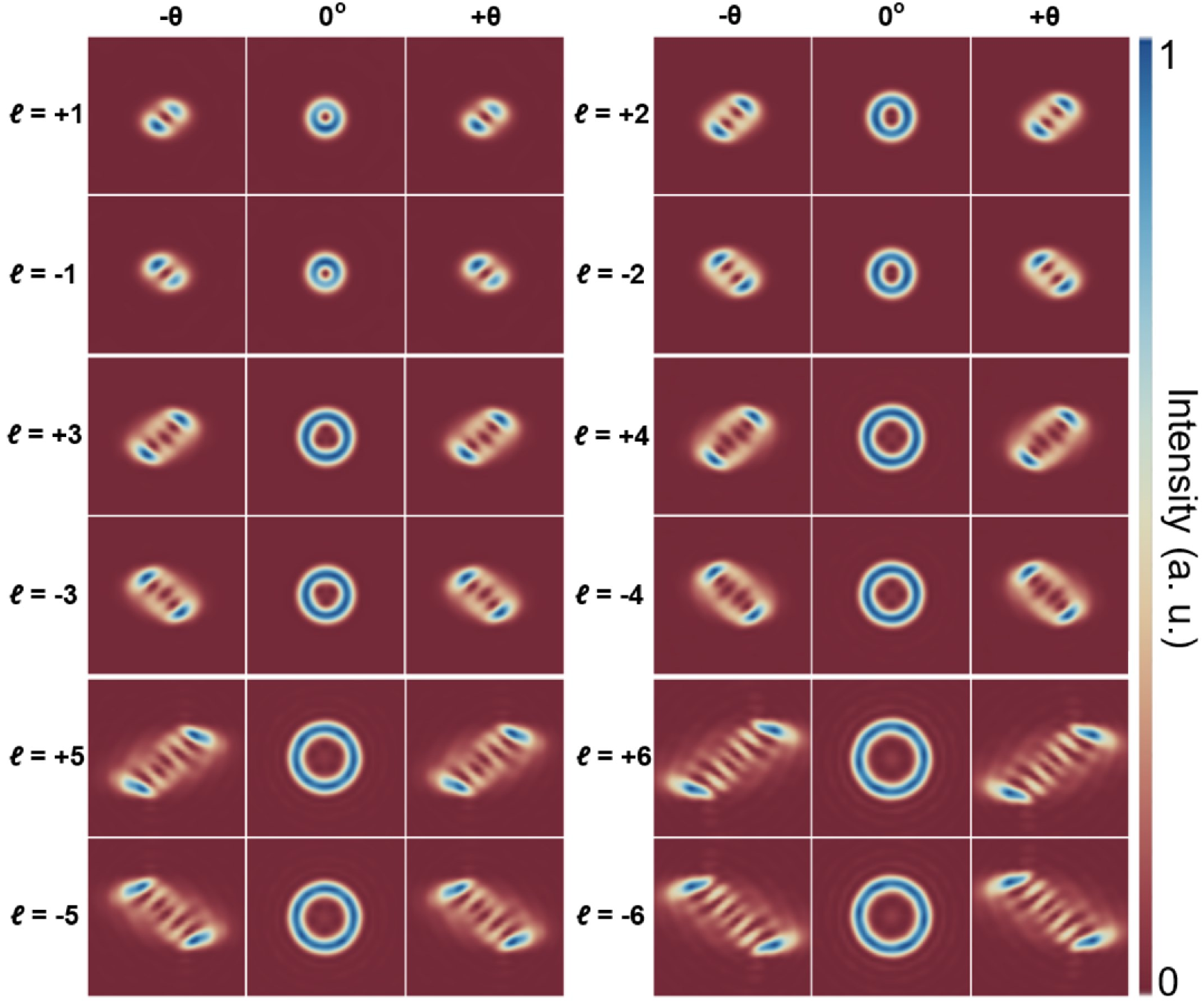}
		\caption{$\mid$ \textbf{Theoretical characterization of the phase topology of OVs.} The distribution profile of the light field of a phase singular beam with topological charge $\ell$ from $\pm1$ to $\pm6$ in the focal plane of the Fresnel lens: under normal incidence of an optical vortex on the NLC lens ($\theta$ = 0 degrees) and when the radiation incidence at angles of $-\theta$ and $+\theta$ ($\theta$ increased from 9 to 18 degrees with increasing topological charge) to the normal of the lens.}
		\label{fig:Supl3}
	\end{figure}

	The pattern of the distribution profile of the light field of a phase singular beam in the direction of -1 the diffraction order (converging beam, $E_{-1}$) during the passage of an optical vortex with a topological charge $\ell$ through a nematic liquid crystal Fresnel lens was analytically calculated using a software system Wolfram Mathematica (Figures \ref{fig:Angle2}b, \ref{fig:focal plane2}a-c, \ref{fig:Supl3}). 
	
	The amplitude of the electromagnetic field $E_{0} (\rho,\phi)$ of an optical vortex incident on the NLC Fresnel lens is represented as a superposition of a Laguerre-Gaussian $E_{LG} (\rho,\phi)$ beam (signal) and Gaussian $E_{G} (\rho)$ beam (noise) in the following form:
	\begin{equation}
		E_{0} (\rho,\phi)=\alpha E_{LG}(\rho,\phi)+\sqrt{1-\alpha^2}E_{G}(\rho),
	\end{equation}
	where $\rho=\sqrt{x^2+y^2}$ and $\phi=\arctan(\frac{y}{x})$ are the radial and angular cylindrical coordinates in the transverse plane of the beam, $x$, $y$, $z$ – coordinates of the point of the singular beam (the origin of the coordinate system is in the center of the beam), $\alpha^2$ – signal intensity amplitude (in simulations $\alpha$ is equal to 0.998), $1-\alpha^2$ – the amplitude of the noise intensity (in simulations equal to 0.4\%).
	
	The Laguerre-Gaussian $E_{LG} (\rho,\phi)$  beam at the beam waist position (e.g. $z=0$) with azimuthal number $\ell$, and radial number $\rho$=0:
	
	\begin{equation}
		E_{LG} (\rho,\phi)=C_{\ell}e^{-\frac{\rho^2}{w_{0}^2}+i\ell\phi}\frac{1}{w_{0}}(\frac{\rho}{w_{0}})^{|\ell|} L_{p}^{|\ell|}(\frac{2\rho^2}{w_{0}^2}),
	\end{equation}
	
	where $L_{\rho}^{\mid\ell\mid}$ is the generalized Laguerre polynomial, $C_{\ell}$ is the normalization constant, and $w_0$ is the beam waist \cite{plick2015physical}.
	
	The amplitude of the Gaussian beam $E_{G} (\rho)$ at $z=0$ with a beam waist equal to $w_{G}$:
	
	\begin{equation}
		E_{G} (\rho)=\frac{e^{-\frac{\rho^2}{w_{G}^2}}\sqrt{2}}{w_{G}\sqrt{\pi}}.
	\end{equation}
	
	The beam amplitude at distance $z$ (Figures \ref{fig:focal plane2}a-c) is simulated by calculating the Fresnel diffraction integral with wavelength $\lambda$=632.8 nm:
	
	\begin{equation}
		E(x,y,z)=\frac{e^{i2\pi z}}{i\lambda z} e^{i\frac{\pi}{\lambda z}(x^2+y^2)}\iint\limits_{-\infty}^{+\infty} E_{0} (x',y')U (x',y')e^{i\frac{\pi}{\lambda z}(x'^2+y'^2)}e^{-i2\pi(\frac{xx'}{\lambda z}+\frac{yy'}{\lambda z})} dx'dy'.
	\end{equation}
	
	The function $U(x', y')$ is the intensity mask of the Fresnel lens with a focal length $f$ = 17.6 cm:
	
	\begin{equation}
		U (x',y')=1-\lfloor \frac{x'^2+y'^2}{\lambda f} mod \ 2\rfloor.
	\end{equation}
	
	\hspace{\parindent}The results of numerical simulation of the distribution profile of an optical vortex diffracted in the -1 direction of the diffraction order (converging beam, $E_{-1}$) with topological charge $\ell$ from  $\pm1$ to $\pm6$ in the focal plane of the Fresnel lens when rotating the NLC element around the vertical axis are shown in the Figure \ref{fig:Supl3}. The simulation results are fully consistent with the experimental results (Figure \ref{fig:All}).
	
\end{document}